\documentclass[aps,reprint,nofootinbib,longbibliography,onecolumn]{revtex4-2}
\usepackage{bm}
\usepackage[latin1]{inputenc}
\usepackage[T1]{fontenc} 
\usepackage{graphicx,multirow}
\usepackage{subcaption}
\usepackage{float}
\usepackage{bm}
\usepackage{braket}
\usepackage{amsmath}
\usepackage{amssymb}
\usepackage{amscd}
\usepackage{latexsym}
\usepackage{slashed}
\usepackage{color, xcolor}
\usepackage{graphicx}
\usepackage[normalem]{ulem}
\definecolor{orange}{cmyk}{0,0.5,1,0}
\usepackage{verbatim}
\usepackage{rotating}
\usepackage{cancel}
\usepackage{caption}
\usepackage{hyperref}
\usepackage{lineno}

\usepackage{makecell,natbib}
\hypersetup{
   colorlinks=true,       
   linkcolor=blue,        
   citecolor=red,         
   filecolor=magenta,      
   }

\begin{document} 

\title{Viability of Boosted Light Dark Matter in a Two-Component Scenario}  

\author{Arindam Basu}
\affiliation{Department of Physics, School of Engineering and Sciences, SRM University-AP, Amaravati, Mangalagiri 522240, India}

\author{Amit Chakraborty}
\affiliation{Department of Physics, School of Engineering and Sciences, SRM University-AP, Amaravati, Mangalagiri 522240, India} 

\author{Nilanjana Kumar,}
\affiliation{Centre for Cosmology and Science Popularization (CCSP), SGT University,
Gurugram, Delhi-NCR, Haryana 122505, India}

\author{Soumya Sadhukhan}
\affiliation{Ramakrishna Mission Residential College (Autonomous) \& Vivekananda Centre for Research, Narendrapur, Kolkata 800103, India } 



\begin{abstract}
We study the boosted dark matter (BDM) scenario in a two-component model. We consider a neutrinophilic two-Higgs doublet model ($\nu$2HDM), which consists of one extra Higgs doublet and a light right-handed neutrino. This model is extended with a light ($\sim 10$~MeV) singlet scalar DM $\phi_3$, which is stabilized under an extra dark $Z_2^{\rm DM}$ symmetry and can only effectively annihilate through the CP even scalar $H$. 
Although the presence of a light scalar $H$ modify the oblique parameters to put tight constraints on the model, the introduction of vectorlike leptons (VLL) can potentially salvage the issue. 
The vectorlike doublet $N$ and singlet $\chi$ are also stabilized through dark $Z_2^{\rm DM}$ symmetry. 
The lightest vectorlike mass eigenstate ($\chi_1 \sim 100$~GeV) is the second DM component of the model. 
Individual scalar and fermionic DM candidates have Higgs/$Z$ mediated annihilation, restricting the fermion DM in a narrow mass region while a somewhat broader mass region is allowed for the scalar DM. However, when two DM sectors are coupled, the annihilation channel $\chi_1 \chi_1 \to \phi_3 \phi_3$ opens up. 
As a result, the fermionic relic density decreases, and paves way for broader fermionic DM mass region with under-abundant relic: 
a region of $[30-65]$ GeV compared to a narrower $[40-50]$ GeV window for the single component case. On the other hand, the light DM $\phi_3$ acquires significant boost from the annihilation of $\chi_1$, causing a dilution in the resonant annihilation of $\phi_3$. This in turn increases the scalar DM relic, allowing for a smaller mass region compared to the individual case. The exact and underabundant relic is achievable in a significant parameter space of the two-component model where the total DM relic is mainly dominated by the fermionic DM contribution. The scalar DM is found to be sub-dominant or equally dominant 
($\sim 30 \% - 80 \%$ of total DM) with significant boost which can be detected in experiments.
\end{abstract}


\maketitle

\setcounter{footnote}{0}




\section{Introduction}
\label{intro}
 
One of the main ignitions behind our quest for new physics is the existence of Dark Matter (DM) whose presence is already established in the experiments, but only through indirect gravitational probes, such as galaxy rotation curves, 
cosmic microwave background radiation and gravitational lensing
\cite{Zwicky:1933gu,Begeman:1991iy,Bertone:2010zza,Bauer:2017qwy,Bertone:2004pz,Lisanti:2016jxe} among others. 
Precise measurements at these experiments
predict the amount of DM in the present Universe to be around ($\sim 26.8\%$) of the total energy density of the Universe. 
In conventional relic density terms, the abundance of DM varies in the range $\Omega h^2 \sim [0.11-0.13]$ at the $3\sigma$ confidence level  \cite{Planck:2015fie}.
The information from the structure formation in the early universe typically prefers the
cold dark matter (CDM) scenario as it fits all the evidence.
If we assume the CDM to be of particle nature~\cite{Bertone:2004pz}, 
then the experiments are yet to
ascertain the exact nature of the dark matter candidate/s. 
Similarly to the visible sector, the dark sector may also consist of multiple dark matter candidates \cite{Zurek:2008qg,Bhattacharya:2019tqq}. 
dark matter candidates can be of different types as well -- fermion, scalar, or gauge boson.

Plethora of dark matter direct and indirect detection experiments are searching for dark matter signal. 
So far, these experiments have only been successful in giving us some hints and limits on dark matter masses and interaction strengths \cite{DAMA:2008jlt,XENON:2020rca,XENON:2022ltv}.
Traditionally, weakly interacting massive particles (WIMPs)~\cite{refId0,Arcadi:2017kky} are proposed as dark matter candidates, 
with masses at the TeV scale, but none of them have been detected so far. 
Hence, the hunt for detecting dark matter in a territory below the GeV scale, which is hitherto not so explored, has begun.
In direct detection experiments, the DM particles collide with the nuclei of the target material and transfer a part of their kinetic energy to
the target. 
In the conventional DM direct detection method, the nuclear recoil energy is then measured by the detectors. 
When dark matter is light, to be precise at the sub-GeV level, the roadblock to their detection
comes from their inability to make the nuclei recoil. This is because the dark matter particles coming from the galaxies
are non-relativistic, with velocity ($v\sim 10^5$ m/s). Hence the kinetic energy of the
recoiled particle is not sufficient to overcome the threshold $\sim \mathcal{O}$ (1 keV-1 MeV) in the direct detection experiments 
such as Xenon \cite{XENON:2020gfr} etc.


{
A light dark matter can only make massive nuclei recoil if it is boosted to relativistic velocities. Examples of boosted DM are those that receive the boost from the cosmic rays \cite{Bringmann:2018cvk,Das:2021lcr,Xia:2022tid,Jho:2021rmn,  Yin:2018yjn}, the diffuse supernova neutrino background (DSNB) \cite{Arguelles:2017atb,Das:2021lcr, DeRomeri:2023ytt}, non-galactic sources \cite{Herrera:2023fpq,Herrera:2021puj},
and blazers \cite{Bhowmick:2022zkj,Wang:2021jic,Granelli:2022ysi,Maity:2022exk}. 
In that case, the light DM can deposit a recoil energy greater than the recoil threshold energy.
This boost not only enhances it detectability in the DM direct detection experiments but also has intricate
connections with the relic density-related phenomenology also. The detection of lower DM masses
is possible only if the scattering cross section is large because the flux of the boosted dark matter is
much smaller than the CDM population in the galaxies.
There have been many studies to address the Boosted Dark Matter (BDM) in the literature (Ref 8-20 of \cite{Bardhan:2022bdg}) where the boost is achieved through different mechanisms. One of the popular mechanisms is when DM inside the galaxy collides with the high-energy cosmic ray particles and gets a boost.

In this paper, we analyze an alternate mechanism following the footsteps of \cite{Agashe:2015xkj}, 
where the heavier dark matter annihilates to light dark matter candidate and the mass difference between them plays a key role to produce the boost. 
We choose this two-component dark matter scenario framework, which is very interesting on its own merit \cite{Belanger:2011ww}.
There are papers in the literature where particular models have been proposed to address the interaction of the boosted dark matter 
via different models mostly in the framework of Self Interacting Dark Matter (SIDM) \cite{Dutta:2021wbn,Borah:2021yek,Baek:2021yos,Kong:2014mia} or others \cite{Ko:2022kvl,Jho:2020sku,Jaeckel:2020oet}.
In such studies dark matter either interacts with the dark gauge boson or portal interactions are assumed. 
Most of the earlier works focus on different external mechanisms to have a boosted DM and their
possible detection intricacies. 
On the other hand, our work concentrates on a different side of the boosted DM sector, focusing on the relic density aspects of the combined two-component DM paradigm, where one of the DM candidate is boosted from the annihilation of the
other relatively heavier DM of the model. Other cosmological aspects of multi-component boosted dark matter scenario is discussed in \cite{Kamada:2021muh}, while the phenomenological aspects of a general multi-component DM scenario is discussed in \cite{Bhattacharya:2022wtr,Bhattacharya:2018cgx,Bandyopadhyay:2023joz}.

We incorporate the boosted dark matter in a two-component model framework, which can be embedded in a larger theory. 
Moreover, this framework not only addresses the relic abundance, but also satisfies other phenomenological constraints 
\cite{Grimus:2007if,Grimus:2008nb,Haber:2010bw} from different experiments such as the LEP, LHC etc. 
In this paper, we have considered a neutrinophilic two Higgs doublet model ($\nu \rm 2HDM$) \cite{Machado:2015sha,Nomura:2017jxb}, 
where one CP even neutral scalar ($H$) can be very light in addition to a light right-handed (RH) neutrinos $N_{R_i}$. 
This model can successfully generate neutrino mass via Type II seesaw mechanism as well as the correct 
relic density, which has been discussed in Ref.~\cite{Mohanty:2018iop}. The $\nu \rm 2HDM$ is further extended with a scalar DM sector to incorporate one 
light gauge singlet scalar $\phi_3$, which is the scalar dark matter candidate in our Model. This is a natural extension as the model already has a MeV scale non-dark scalar which can play the role of a mediator effectively in a DM annihilation, only when DM mass is also at the same scale. Possibility of a light DM is a novel extension of the neutrinophilic 2HDM. We also extend the model with vectorlike fermions. Addition of the vectorlike fermions 
do allow for more favorable parameter space when precision observable constrains are considered, see \cite{Ellis:2014dza} for example. We choose one vectorlike doublet ($N$) and one vectorlike singlet ($\chi$) extending the $\nu \rm 2HDM$ framework and we obtain the fermionic DM $\chi_1$ from the mixing of the former two vectorlike particles.

In this two-component DM model, the boosting is achieved through annihilation $\chi_1\chi_1\to \phi_3 \phi_3$.
This is the process that couples the vectorlike lepton and the scalar dark sector. 
The relic density of the dark matter is achieved by solving the Boltzmann Equations (BE) for two cases:
I. When the scalar and vectorlike sectors are uncoupled, and II. when these two sectors are coupled. 
We study both scenarios in detail and present a comparison of the 
individual and coupled scenario of the DM. 
In both the scenarios, we obtain the correct relic density for some selected benchmark points. 
We also show that scalar DM gets a sufficient boost in the coupled scenario, leading to significant modification in its DM phenomenology. 
Moreover, unlike other scenarios studied in the literature, both DM particles interact with the SM giving a very rich phenomenology.
Overall, we present an well established Beyond Standard Model scenario to arrange a complete two-component dark matter model, boosted DM being important feature of it.
}

%
%
\section{Extended Neutrinophilic Two-Higgs doublet model}
\label{numodel}

{We consider the neutrinophilic Two-Higgs doublet model, in-short $\nu$2HDM \cite{Machado:2015sha,Nomura:2017jxb}, extended with an additional scalar singlet $\phi_3$ which is the scalar dark matter candidate, and electroweak gauge singlet right-handed (RH) neutrinos, $N_{R_{i}}$ ($i$=1,2,3), for each flavor of Standard Model (SM) leptons.    
%
The $\nu$2HDM theory is based on the symmetry group $SU(3)_C \times SU(2)_L \times U(1)_Y \times Z_2$, and include two Higgs doublets, $\Phi_{1}$ and $\Phi_{2}$. 
%
All charged fermions of SM and the Higgs doublet $\Phi_{1}$, are even under the discrete symmetry, $Z_2$, while RH neutrinos and the Higgs doublet $\Phi_{2}$ are odd under $Z_2$. The right-handed neutrinos help to solve the neutrino mass problem, discussed in detail below. The particle/field content of the model introduced until now is listed in Table~\ref{table:1}.}\\
\begin{table}[h!]
\centering
\begin{tabular}{ |c|c|c|c|c|c| } 
\hline
Particle Name & $SU(2)_L$ Charges &  $U(1)_Y$ Charges & $Z_2$  Charges & $Z_2^{\rm DM}$ Charges\\
\hline
\multicolumn{5}{| c |}{\bf Scalar Fields}\\
\hline
 $\Phi_1$ & 2 & 1 &  1 & 1\\
\hline 
$\Phi_2$ &  2 & 1  & -1 & 1 \\ 
\hline
$\phi_3$ & 1 &  0  &  1 & -1 \\ 
\hline 
\multicolumn{5}{| c |}{\bf Fermionic Fields}\\
\hline
$N_{R_{i}}$ & 1 & 0  &  -1  & 1\\
\hline
\end{tabular}
\caption{The BSM fields and their charge assignments.}
\label{table:1}
\end{table}

\newpage
\begin{itemize}
\item \underline{Neutrinophilic Two-Higgs doublet model}: \\
The most general scalar potential for a CP-conserving 2HDM \cite{Branco:2011iw} can be written as, 
\begin{eqnarray}
 V &=&m_{11}^2\Phi_{1}^{\dag}\Phi_{1}+m_{22}^2\Phi_{2}^{\dag}\Phi_{2}- (m_{12}^2~\Phi_{1}^{\dag}\Phi_{2}+h.c.)+ \frac{\lambda_1}{2}(\Phi_{1}^{\dag}\Phi_{1})^{2}+\frac{\lambda_2}{2}(\Phi_{2}^{\dag}\Phi_{2})^{2} \nonumber \\ 
 &+&\lambda_3(\Phi_{1}^{\dag}\Phi_{1})(\Phi_{2}^{\dag}\Phi_{2}) + \lambda_4(\Phi_{1}^{\dag}\Phi_{2})(\Phi_{2}^{\dag}\Phi_{1}) \nonumber \\
 &+& \Bigg[\frac{\lambda_5}{2}(\Phi_{1}^{\dag}\Phi_{2})^2+ 
\lambda_6 (\Phi_1^\dag \Phi_1)(\Phi_1^\dag \Phi_2) + \lambda_7 (\Phi_2^\dag \Phi_2)(\Phi_1^\dag \Phi_2) + h.c.\Bigg].
\label{eq:1}
\end{eqnarray}
where $\Phi_1$ and $\Phi_2$ are two $SU(2)_L$ Higgs doublets. If we assume a $Z_2$ symmetry under which 
$\Phi_1$ is even and $\Phi_2$ is odd, then the coefficients $m_{12}$ = $\lambda_6$ = $\lambda_7$ = 0. However, one can 
allow a softly broken $Z_2$ symmetry assuming small non-zero values of $m_{12}$ while keeping $\lambda_6$ = $\lambda_7$ = 0. 

The elements of the two Higgs doublets are in general complex and can be written (in unitary gauge) as,
\begin{align}
\Phi_{1} =  \frac{1}{\sqrt{2}} \left(\begin{array}{c}
                                                 \sqrt{2} (v_2/v)H^{+} \\
                                                 h_0 + v_1 + i (v_2/v)A  \\
                                               \end{array}
                                             \right),    \nonumber
\end{align}
\begin{align}
           \Phi_{2} = \frac{1}{\sqrt{2}} \left(
                                               \begin{array}{c}
                                                 -\sqrt{2} (v_1/v) H^{+} \\
                                                 H_0 + v_2- i (v_1/v) A \\
                                               \end{array}
                                             \right) \label{eq:2}
\end{align} 
where $v_1$ and $v_2$ are the vacuum expectation values (vev) of the two Higgs 
doublets, namely $v_1 = \langle\Phi_{1}\rangle$ and $v_2 = \langle\Phi_{2}\rangle$ 
with $v^2 = v^2_{1} + v^2_{2}$. 
After the electroweak symmetry breaking (EWSB), the two Higgs doublets mix with each other to 
produce two CP-even neutral Higgs bosons (h and H), one CP-odd neutral Higgs boson (A) and a pair 
of charged Higgs boson ($H^{\pm}$). These neutral scalars provide the portal interaction between the SM and DM particles. Three massless Goldstone bosons are also generated which resulted 
into the longitudinal modes of the W and Z bosons. The mass eigenstates h, and H are related to the weak eigenstates $h_0$, and $H_0$ by, 
\begin{eqnarray} 
h_0 = H \cos\alpha - h \sin\alpha, ~H_0 = H \sin\alpha + h \cos\alpha.
\end{eqnarray} 
There is an orthogonal mixing between the charged and CP-odd Higgs states 
with corresponding charged and neutral Goldstone modes with a mixing angle $\beta$, where 
$\tan \beta = \frac{v_2}{v_1}$. 



The masses of physical Higgs bosons (both charged and neutral) can be written as,
\begin{eqnarray} \nonumber
    m_h^2&=&M^2c_{\beta-\alpha}^2+v^2\Big[\lambda_1 s_\alpha^2c_\beta^2+\lambda_2 c_\alpha^2 s_\beta^2-\frac{1}{2}\lambda_{345}~\sin2\alpha~ \sin2\beta\Big] \\ \nonumber
    m_H^2&=& M^2s_{\beta-\alpha}^2+v^2\Big[\lambda_1 c_\alpha^2c_\beta^2+\lambda_2 s_\alpha^2 s_\beta^2+\frac{1}{2}\lambda_{345}~\sin2\alpha~ \sin2\beta\Big] \\ \nonumber
    m_A^2&=& M^2-\lambda_5~v^2\\ 
    m_{H^\pm}^2&=& M^2-\frac{\lambda_{45}}{2}~v^2 \label{eq:higgs-mass}
\end{eqnarray}
where $\lambda_{345}=\lambda_3 + \lambda_4 + \lambda_5$,\quad $\lambda_{45}=\lambda_4+\lambda_5$, 
$M^2=\frac{m_{12}^2}{s_\beta~c_\beta}$, and 
\begin{eqnarray}
    \tan2\alpha=\frac{-m_{12}^2+\lambda_{345}~v^2\sin2\beta}{-2m_{12}^2~\cot2\beta+v^2(\lambda_1~c_\beta^2-\lambda_2~s_\beta^2)+v^2~\lambda_{345}\cos2\beta}.
\end{eqnarray} 

One can always choose to work on the basis of physical mass over the gauge basis and assume that the physical masses are
the free parameters along with the Higgs mixing angle and $v_2$. In this basis, the quartic couplings
can be expressed in terms of the physical masses, vevs, and mixing angles. 
We ensure that all the quartic couplings satisfy the limits coming from the stability of the vacuum and tree-level perturbative unitarity.

The triple Higgs coupling constants, in terms of the free parameters of the model, can be written as, 
\begin{eqnarray}
\lambda_{hHH}&=&\frac{1}{2}(\lambda_3+\lambda_4+\lambda_5)\Big[v_2c_\alpha^3-v_1s_\alpha^3+2v_1s_\alpha c_\alpha^2-2v_2s_\alpha^2c_\alpha\Big] \nonumber \\
&+&\Big[3\lambda_2v_2c_\alpha s_\alpha^2-3\lambda_1 v_1s_\alpha c_\alpha^2\Big]  \label{eq:2.20} \\
\lambda_{hhH}&=&\frac{1}{2}(\lambda_3+\lambda_4+\lambda_5)\Big[v_2s_\alpha^3+ v_1c_\alpha^3-2v_2s_\alpha c_\alpha^2-2v_1s_\alpha^2c_\alpha\Big] \nonumber \\
&+& \Big[3\lambda_1v_1c_\alpha s_\alpha^2+3\lambda_2 v_2s_\alpha c_\alpha^2\Big]   \label{eq:2.21} \\
\lambda_{hhh} &=& (\lambda_3+\lambda_4+\lambda_5)\Big[v_2c_\alpha s_\alpha^2-v_1s_\alpha c_\alpha^2\Big]+\Big[\lambda_2v_2c_\alpha^3 - \lambda_1 v_1 s_\alpha^3\Big]  \label{eq:2.22} \\ 
 \lambda_{HHH} &=&(\lambda_3+\lambda_4+\lambda_5)\Big[v_2c_\alpha^2 s_\alpha+v_1s_\alpha^2 c_\alpha\Big]+\Big[\lambda_2v_2s_\alpha^3 + \lambda_1 v_1 c_\alpha^3\Big] \label{eq:2.23}
\end{eqnarray}

\item {\underline{Fermion sector: Right-handed Majorana Neutrinos}}: \\
{ 
To generate the neutrino mass in this model, we introduce three right handed fermions, that are gauge singlet Majorana neutrinos 
$N_{R}$, all of which transform as odd under the $Z_2$ symmetry, while all the SM fermions being $Z_2$ invariant. 
The Yukawa Lagrangian in the flavor basis takes the form,
\begin{equation}
\mathcal{L}_{Y}= Y^{d}_{\alpha \beta}\bar{Q}_{L,\alpha} \Phi_{1} d_{R,\beta} + Y^{u}_{\alpha \beta}\bar{Q}_{L,\alpha} \tilde{\Phi}_{1} u_{R,\beta} + Y^{l}_{\alpha \beta} \bar{L}_{L,\alpha}\Phi_{1}l_{R,\beta} + Y^{\nu}_{\alpha \beta} \bar{L}_{L,\alpha} \tilde{\Phi}_{2} N_{R,\beta}+ \mathrm{h.c.}
\end{equation}
where the doublets are $\bar{Q}_L^T=(u_L \ \ d_L)$ and $\bar{L}_L^T=(\nu_L \ \ l_L)$. 
Here the neutrino mass matrix is a $6 \times 6$ matrix with $Y^{\nu}_{\alpha \beta}$ providing the off-diagonal entries of the $3 \times 3$ Yukuwa matrix. The small vacuum expectation value of the second Higgs doublet, allows us to achieve a low scale seesaw mechanism in this case, which was first proposed in Ref.~\cite{Ma:2000cc}.

%

%

Apart from the neutrino mass generation mechanism, to account for the limits on neutrino oscillation parameters, namely the two mass square differences, three mixing angles, and one CP phase, the new physics model must include a minimum of two right-handed fermions. Introducing three right-handed neutrinos and following the Casas-Ibarra parametrization, it was shown that the neutrino Yukawas can be expressed 
as \cite{Huitu:2017vye},
$$ Y^{\nu} = \frac{1}{v_2} \sqrt{M_{R}^{\rm diag}} R \sqrt{M_{\nu}^{\rm diag}} U^{T}, $$
%
where $U$ is the standard PMNS matrix, $R$ is an orthogonal matrix that can be taken as the unit matrix, $M_R$ is the right-handed neutrino mass matrix and $M_{\nu}$ is the neutrino mass matrix that produces a neutrino spectrum that satisfies the oscillation data. In our set-up, one of the right-handed neutrino masses is pushed down to as low as of 10 MeV, which is cosmologically viable as discussed in Refs.~\cite{Nollett:2014lwa,Blennow:2019fhy,Escudero:2018mvt}, while others can be somewhat heavier. 
%
%

After the EW symmetry breaking, the couplings of the Higgs bosons $h$ and $H$ with the right-handed neutrino are given by, 
\begin{eqnarray}
\lambda_{hN_R\nu_L}=-\frac{y_\nu}{\sqrt{2}}s_\beta ~~; ~~
\lambda_{HN_R\nu_L}=\frac{y_\nu}{\sqrt{2}}c_\beta.
\end{eqnarray}
}

%

%
%
%

 
\item {\underline{Scalar Dark Matter sector}}: \\
As stated above, we extend the model to include a gauge-singlet scalar $\phi_3$ that is odd under an extra stabilization symmetry $Z_2^{\rm DM}$. The scalar potential now contains the following additional terms in terms of the scalar DM field $\phi_3$,
\begin{equation}
V_{DM}=\frac{1}{2}\mu_{\phi_3}^{2}\phi_{3}^{2}+\frac{\lambda_{\phi_3}}{4!} \phi_{3}^{4}+\kappa_1 \Phi_{1}^{\dag}\Phi_{1} \phi_{3}^{2} + \kappa_2 \Phi_{2}^{\dag}\Phi_{2} \phi_{3}^{2}.
\end{equation}
This potential gives the physical mass of the scalar DM $\phi_3$ as,
\begin{equation}
    m_{\phi_3}^2 = \mu_{\phi_3}^2+\kappa_1 v_1^2 +\kappa_2 v_2^2
\end{equation}
The above potential includes four free parameters, namely $\mu_{\phi_3}^2$, $\lambda_{\phi_3}$, $\kappa_1$ and $\kappa_2$, 
among these $\lambda_{\phi_3}$ does not play any role for Higgs-portal interactions, so we set this to unity. 
The couplings between the DM particle ($\phi_3$) and the other neutral scalars play a crucial role in the production of the light DM particle ($\phi_3$).
In the alignment limit $c_{\beta-\alpha}=0$, these portal couplings can be expressed as:
\begin{eqnarray}
\lambda_{\phi_3\phi_3 h}=(\kappa_1c_\beta^2+\kappa_2s_\beta^2)v \\
\lambda_{\phi_3\phi_3 H}=\frac{1}{2}(\kappa_1-\kappa_2)\sin2\beta~v \\
\lambda_{\phi_3\phi_3hh}=\frac{1}{2}(\kappa_1c_\beta^2+\kappa_2s_\beta^2) \\
\lambda_{\phi_3\phi_3HH}=\frac{1}{2}(\kappa_1s_\beta^2+\kappa_2c_\beta^2)\\
\lambda_{\phi_3\phi_3hH}=\frac{1}{2}(\kappa_1-\kappa_2)\sin2\beta.
\end{eqnarray}


\end{itemize}
At this point, let us discuss the mass spectrum of the particles introduced till now. We choose the mass parameter $\mu_{\phi_3}^2$ in such a manner that the physical mass of the scalar DM 
particle lies in the MeV scale (or at maximum up to a GeV), which in turn helps us to incorporate a light mediator particle in the model\footnote{We refer \cite{Binder:2022pmf} for a global analysis of resonance-enhanced light 
scalar dark matter with a mass of 0.3-2 GeV.}. To achieve the small mediator mass, we set $v_1$ = 246 GeV and vary $v_2$ between 1 keV and 100 MeV, so that the alignment limit is satisfied, that is, $\beta - \alpha \approx \frac{\pi}{2}$. This results in very small $\tan\beta \sim$ $\mathcal{O}(10^{-6})$,
which in turn gives very small values of $ \cos(\beta - \alpha)$. Hence, the neutral scalar $h$ effectively behaves like an SM Higgs boson at 125 GeV and a very light CP-even neutral Higgs boson ($H$) mediator with mass around a few MeV is obtained.




\section{Theoretical and Experimental Constraints, and the vectorlike Lepton} 
{In this section, we explore different constraints on the extended $\nu$2HDM scenario, to probe its viability for the parameter region of interest. These constraints include theoretical constraints, experimental bounds on BSM particles, electroweak precision test, among others. We observe that with the given particle spectrum, the oblique parameter S moves towards large negative values, mainly aided by the contribution from the light (MeV range) scalar H, which is in tension with the experimental data. To resolve this, we augment the existing scenario with vectorlike leptons (VLL), which also act as another dark matter candidate along with the existing scalar DM $\phi_3$. }
\subsection{Constraints and Direct Search Limits}
\begin{itemize}
\item \underline{Theoretical constraints}: \\ 
The stability and perturbative unitarity conditions at tree level will constrain some of the parameters of the scalar potential. These constraints will eventually restrict the range of physical scalar masses and their couplings. To have stability at
tree-level, the following constraints should be satisfied, 
\begin{equation}
    \lambda_{1,2} >0, \quad \lambda_3 > -\sqrt{\lambda_1 \lambda_2}, \quad \rm{and} \quad \lambda_3 + \lambda_4 - |\lambda_5| > -\sqrt{\lambda_1 \lambda_2}. \label{eq:stability-const}
\end{equation}
If the quartic couplings are excessively large, the leading-order amplitudes for scalar-scalar scattering could violate unitarity at high energy scales, demanding additional physics to resolve this issue. Hence, the following conditions are imposed on the quartic couplings,
\begin{eqnarray}
    |\lambda_3\pm\lambda_4|&\leq&8\pi\\
    |\lambda_3\pm\lambda_5|&\leq&8\pi\\
    |\lambda_3+2\lambda_4\pm3\lambda_5|&\leq&8\pi\\
    \Big|\frac{1}{2}\Big(\lambda_1+\lambda_2\pm\sqrt{(\lambda_1-\lambda_2)^2+4\lambda_4^2}\Big)\Big|&\leq&8\pi\\
    \Big|\frac{1}{2}\Big(\lambda_1+\lambda_2\pm\sqrt{(\lambda_1-\lambda_2)^2+4\lambda_5^2}\Big)\Big|&\leq&8\pi\\
    \Big|\frac{1}{2}\Big(3\lambda_1+3\lambda_2\pm\sqrt{9(\lambda_1-\lambda_2)^2+4(2\lambda_3+\lambda_4)^2}\Big)\Big|&\leq&8\pi
\end{eqnarray} 
Some authors may prefer to impose a somewhat stronger limit of 4$\pi$, however, we have checked that the results remain the same even after considering the stronger limit. An important point to note here: the parameter $m_{12}^2$ plays a crucial role in determining the minima of the scalar potential. In our work, we choose $m_{12}^2 = 0$ satisfying exact $Z_2$ symmetry, which implies $v_1 \sim v$ and $v_2$ takes small values for $\tan\beta<<1$ and very small values of $\cos_{\beta-\alpha}$ in the alignment limit. In this case, one of the scalars behaves almost like the SM Higgs boson while the other one as inert.  
%
%
%
\item \underline{Bounds from Neutral and Charged Higgs boson searches}: \\
We expect that the constraint on the various couplings of the SM Higgs boson with the SM particles is trivially satisfied by $h$. The choice of alignment limit, and thereby small mixing among the Higgs doublets drives this expectation. However, the existence of a light mediator particle ($H$) and a light scalar DM candidate leads to an additional contribution to the SM Higgs boson width through the invisible mode $h \to H H$ and $h \to \phi_{3} \phi_{3}$. 
The small values of $\tan\beta$ and $c_{\beta-\alpha}\approx 0$ approximate the triple Higgs coupling $\lambda_{hHH}$ as follows:
\begin{eqnarray}
    \lambda_{hHH}\approx \frac{1}{2}\lambda_{345}~v_1~c_{\beta}^3 \label{eq:3.8}
\end{eqnarray}
where, the $\lambda_{345}=\lambda_3+\lambda_4+\lambda_5=\frac{1}{v^2}\Big[\frac{m_{12}^2}{c_\beta~s_\beta}+\frac{\sin2\alpha}{\sin2\beta}(m_h^2-m_H^2)\Big]$. Depending on the choices of the free parameter, $\lambda_{hHH}$ can take large values, which in turn would contribute to a large Higgs invisible decay width. In this study, we assume that the total rate at which the SM-like Higgs boson (h) decays invisibly is less than 15\%, which is consistent with the measurements \cite{CMS:2023sdw}. In fact, using {\tt HiggsTools} \cite{Bahl:2022igd}, we check that the choices of the masses and couplings used for the benchmark point for later sections are consistent with the measurements of the SM-like Higgs boson ($h$) at the LHC. {The lightest neutral Higgs boson H decays predominantly to neutrinos, therefore leading to the final state with missing transverse energy only. These particles will not receive any strong bound from direct searches at the LHC, other than the strong constraint from the measurement of the invisible width of the SM Higgs boson, which has been considered already in our analysis. }\\ \\
{
The charged Higgs production at the LHC in the context of $\nu$2HDM is same as in 2HDM, however, due to the smallness of the mixing between the two Higgs bosons, the decays of charged Higgs to quarks are highly suppressed by the mixing factor $\tan \beta$. Charged scalars decaying through $H^{\pm}\rightarrow l^{\pm} \nu$ modes were already searched at the LEP and the mass of the charged Higgs was constrained as $m_{H^{\pm}}$> 80 GeV \cite{ALEPH:2013htx}. 
The smallness of $v_2$ and given the mass spectrum, the dominant decay mode of $H^{\pm}$ is $H^{\pm}\rightarrow W^{\pm} H$ with $H$ decaying to neutrinos. Evidently, search for charged Higgs boson through this mode will receive large irreducible SM backgrounds. Note that charged Higgs bosons can also contribute to the rate of SM-like Higgs boson decay into di-photon through loops \cite{Seto:2015rma}. The pseudoscalar Higgs $A$, on the other hand, couples dominantly to neutrinos. Therefore, we set the masses of the charged Higgs boson and the CP-odd neutral Higgs boson at 500 GeV, so that all existing collider bounds are satisfied.} \\

\item \underline{Invisible width of Z boson}: \\
The invisible width $Z$ comes from the decays $Z\to S \nu_L \bar{\nu_L}$ with $S=\phi_3, A, H$ and $m_S<m_Z$. The model does not possess the neutral Higgs scalar Higgs $A$ with a mass less than $m_Z$. Rather, we have set the mass $m_A=m_{H^\pm}=500~{\rm GeV}$. Therefore, in this scenario, we can discard the discussion on the constraints of the invisible width $Z$ for $Z\to A\nu_L\bar{\nu_L}$. And for the MeV scale scalar DM $\phi_3$, and the very light neutral scalar Higgs $H$, the $Z$ invisible width $\Gamma_S$ and $Z\to H\nu_L\bar{\nu_L}$ does not contain significant sensitivity due to the suppressed coupling with the $Z$ boson. Therefore, no additional constraints are expected to come from the invisible width of the $Z$ boson as the current bound will be trivially satisfied \cite{CMS:2022ett}.  
{
\item \underline{Neutrino oscillation data:}\\
In the neutrinophilic Higgs doublet model ($\nu$HDM), an additional Higgs doublet with a suppressed vacuum expectation value (vev) is introduced to facilitate the generation of small neutrino masses via the seesaw mechanism. This framework allows for sizable Dirac neutrino Yukawa couplings, which have significant phenomenological implications. A comprehensive analysis of neutrino oscillation data within this model reveals a relationship between the Yukawa coupling $y_{\nu}$, the vev $v_2$, and the RHN mass $m_{N_R}$ as detailed in \cite{Huitu:2017vye}.
\begin{equation}
    |y_{\nu}(v_2,m_{N_R})| \approx 0.63 \times \frac{100 ~\rm keV}{v_2} \sqrt{\frac{m_{N_R}}{100~ \rm GeV}} \label{eq:Ynu-v2}
\end{equation}
\begin{figure}[h!]
    \centering
    \includegraphics[width=0.4\linewidth]{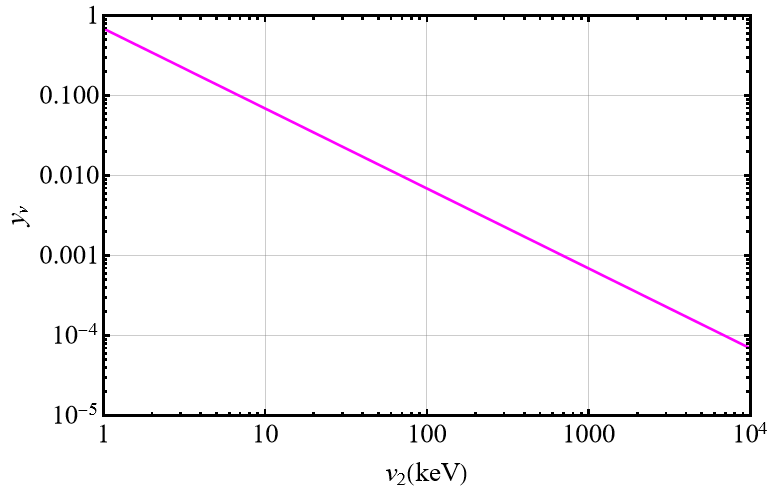}
    \caption{{The variation of  the Yukawa coupling $y_{\nu}$ with vev $v_2$ for $m_{N_R}=$ 10 MeV, satisfying the neutrino oscillation data. }}
    \label{fig:Ynu-v2}
\end{figure}
Following, the Eq. \ref{eq:Ynu-v2}, the Fig. \ref{fig:Ynu-v2}, shows the Yukawa couplings for the range of vev $v_2$ at a fixed RHN mass 10 MeV. 
To provide a rough estimate: with $v_2 \approx 2$~ MeV and the mass of right-handed neutrinos $m_{N_R} \sim 10$~ MeV, one can generate the correct masses and mixings of Majorana neutrinos when the elements of the Yukawa matrices $Y^{\nu}$ take up to a maximum value $y_{\nu} \sim O(10^{-4})$. For a more detailed discussion on the neutrino oscillation data in $\nu$ 2HDM, see~\cite{Cherchiglia:2023utd,Huitu:2017vye}.
}
{
\item \underline{Anomalous magnetic moments and Flavor data:} \\
The charged scalars could also have an impact on the charged lepton $g-2$, but the associated one-loop amplitude is suppressed. Our calculations show that the one-loop contributions to both the muon and electron $g-2$ are negligible because of this suppression. Furthermore, the precision in measuring $(g-2)_{\tau}$ is insufficient to impose a meaningful constraint. In a general 2HDM, it has been observed that two-loop Barr-Zee diagrams can potentially exceed the contributions from one-loop diagrams. However, this is not the case in neutrinophilic 2HDM, as the charged lepton couplings to the charged Higgs ($H^\pm$) and pseudo-scalar ($A$) are suppressed by a factor of $\tan\beta$. Consequently, we conclude that the measurements of the electron, muon, and tau $g-2$ do not impose any constraints on this scenario.\\
Flavor-violating decays involving quarks, such as $b \to s \gamma$ and $B_s \to \mu^+ \mu^-$, do not receive additional contributions from BSM particles because their couplings to SM fermions are strongly suppressed. Note that charged scalars can mediate lepton flavor-violating decays, such as $\mu \to e \gamma$ and $\mu \to 3e$. However, the corresponding branching ratios scale inversely with the fourth power of the mass of the charged scalars, $m_{H^\pm}$, which means that choosing $m_{H^\pm} = 500$~ GeV ensures that these bounds are also satisfied. Several other astrophysical observations will also be easily satisfied; for details, we refer to \cite{Bertuzzo:2015ada, Sher:2011mx}.} 
%
%
%
%
\item \underline{Electroweak precision data (Oblique Parameters)}: \\
The oblique parameters ($S,T,U$) \cite{Peskin:1991sw} play a crucial role in constraining the parameter space of any given particle physics model having relatively light particles \cite{Grimus:2007if,Grimus:2008nb,Haber:2010bw}.
In particular, the $S$ parameter is highly sensitive to the
effects of new physics at scales lower than $M_Z$. More precisely, the $S$ parameter captures the evolution of the two-point neutral gauge boson two-point functions ($ZZ,~Z\gamma,~\gamma \gamma$) from zero momentum. 
In the context of $\nu \rm 2HDM$, which includes a MeV-scale neutral scalar Higgs boson (H), the $S$ parameter places significant constraints on the model. The T parameter, on the other hand, quantifies the degree of custodial symmetry breaking at zero momentum and restricts the parameter space for particles charged under $SU(2)_L$. Similarly, the $U$ parameter reflects corrections to the $W$ boson mass introduced by light-charged scalar particles in loop processes. In the $Z_2$ symmetric $\nu \rm 2HDM$, the structure of quartic couplings with small values $\tan \beta$ in the alignment limit results in a SM-like Higgs boson accompanied by a CP even neutrinophilic neutral scalar H of mass in the MeV scale. Hence, it becomes highly sensitive to Electroweak Precision Test (EWPT) observables. The choice of a charged Higgs mass at 500 GeV helps to overcome the constraints coming from the U-parameter. However, it is important to check the status for the $S$ and $T$ parameters. 
\\
We use \texttt{ 2HDMC} \cite{Eriksson:2009ws}, a publicly available code, designed for general \textit{CP} conserving 2HDM, to calculate the oblique parameters for our model. To be precise, we follow the methodology for calculating the oblique parameters in $\nu \rm 2HDM$ as obtained in \cite{Haber:2010bw}. For the bound on the oblique parameters, we use the latest {\tt GFITTER} values for the best fit and associated uncertainties \cite{Baak:2014ora}, assuming the reference mass of the Higgs boson and Top quark at 125 GeV and 173 GeV respectively, 
\begin{equation}
   \Delta S^{\rm SM}=0.05\pm 0.11, \quad  \Delta T^{\rm SM}=0.09\pm 0.13, \quad {\rm and} \quad \Delta U^{\rm SM}=0.01\pm 0.11. \label{eq:STU-numbers}
\end{equation}
%
%
To reassure ourselves about the different constraints (both theoretical and experimental) imposed on the parameter space under consideration, we perform a dedicated scan over the physical parameter space satisfying the theoretical constraints and current bounds, including the measurements of the invisible decay width of $Z$ and the SM Higgs boson. After setting mass of the SM-like Higgs boson at 125 GeV, $v_1$ = 246 GeV, $m_A = m_{H^\pm} = 500$ GeV\footnote{This particular choice is motivated by the observed limit coming from T-parameter. }, we scan over the parameters $v_2$, $m_H$, $\alpha$, $\mu_{\phi_3}^2$, $\kappa_1$, and $\kappa_2$ in the following ranges, 
\begin{eqnarray} \label{eq:param-scan}
{1 ~{\rm keV} < v_2 <100 ~{\rm MeV}}, \quad 1 ~{\rm MeV}< m_H <100 ~{\rm MeV}, \quad -\pi/2 < \alpha <\pi/2, \\ \nonumber 
-1500~{\rm GeV^2} < \mu_{\phi_3}^2  < 0 ~ {\rm GeV^2}, \quad 0.001 <  \kappa_1  <0.02, \quad 0.001 <  \kappa_2  <0.02. 
\end{eqnarray}

We compute the oblique parameters for all the points allowed by the remaining theoretical and experimental constraints. In Figure \ref{fig:scan1}, we show the available parameter space after imposing all the theoretical constraints (except the oblique parameters) as well as the direct search bounds including the measurements of the invisible decay widths of Z and the SM Higgs boson. The blue points are (randomly) sampled points following the parameter ranges mentioned above (Eq. \ref{eq:param-scan}), while the red points are the ones allowed by the constraints.

\vskip 0.2cm
The key observations from the parameter space scan are listed below: 
\begin{itemize}
\item The values of the Higgs mixing angle $\alpha$ and $\tan\beta$ are such that $\cos (\beta-\alpha)$ is very small (in fact $\cos{(\beta-\alpha)}<10^{-4}$), as shown in the left plot of Figure \ref{fig:scan1}. {This is desirable as it brings the properties of scalar h to be aligned with the SM Higgs boson.
\item Following Eq. \ref{eq:higgs-mass}, the light mediator Higgs boson $H$ mass has a linear dependence on the vev of the second Higgs doublet, i.e $v_2$. By imposing the perturbative unitarity conditions on the randomly sampled points, we get $m_H \lesssim 10\times v_2$ as shown by the red points. A light mediator naturally helps in probing low mass scalar DM, therefore we fix $m_H$ at 20~MeV. This requires the minimum of vev $v_2$ to be approximately 2 MeV. Consequently, the Yukawa coupling $y_{\nu}$ must not exceed $4 \times 10^{-4}$ as shown in Fig. \ref{fig:Ynu-v2}.}  
\item The two most important trilinear couplings involving scalars that contribute to the invisible widths of Z and SM Higgs are $\lambda_{hHH}$ and $\lambda_{\phi_3 \phi_3 h}$.  We find that all allowed points have $\lambda_{hHH} < 1.5$ GeV, while $\lambda_{\phi_3 \phi_3 h}$ can vary over a wider range up to $\lambda_{\phi_3 \phi_3 h} < 3$ GeV.  
\item The best possible value of $\kappa_1$ is found to be 0.01, while we find that the results remain almost insensitive to $\kappa_2$. Hence, we set $\kappa_2$ to 0.002. These choices of $\kappa_1$ and $\kappa_2$ are used for the rest of our analysis.
\end{itemize}
\begin{figure}[htb!]
    \centering
   \includegraphics[width=0.3\linewidth]{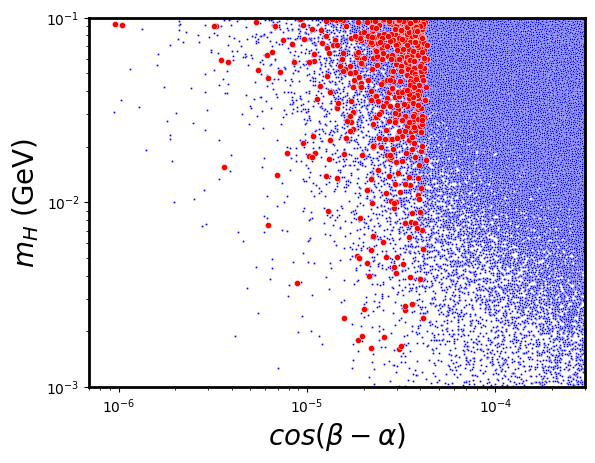}
   \includegraphics[width=0.3\linewidth]{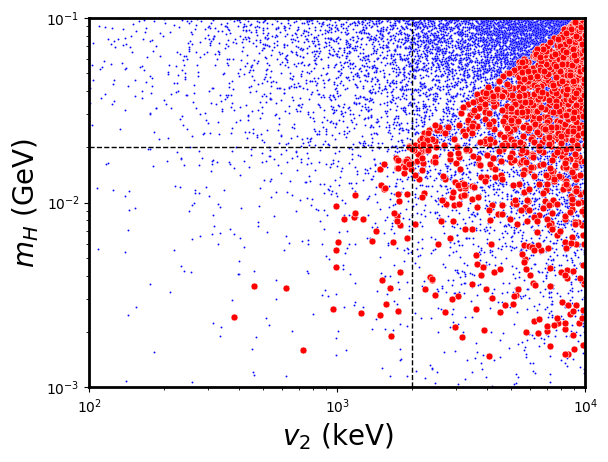}
   \includegraphics[width=0.3\linewidth]{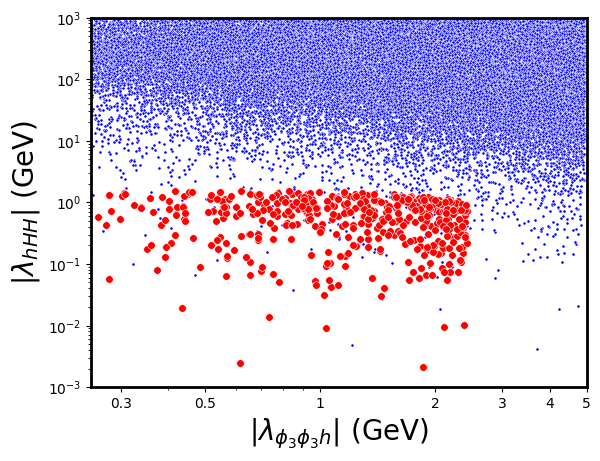} 
    \caption{The parameter space points in the planes of different observables, here the blue points are obtained through a parameter space scan (Eq. \ref{eq:param-scan}), while the red points represent the points allowed by theoretical and experimental constraints except the oblique parameters.}
    \label{fig:scan1}
\end{figure}
In Figure \ref{fig:STU}, we present the values of the oblique parameters S and T for all points allowed by the previously mentioned constraints. It is clear from the figure that the red points are largely disfavored by the data (see Equation \ref{eq:STU-numbers}). Our findings are consistent with previous studies, such as those in \cite{Machado:2015sha,Mohanty:2018iop}. 
\begin{figure}[htb!]
    \centering
    \includegraphics[width=0.5\linewidth]{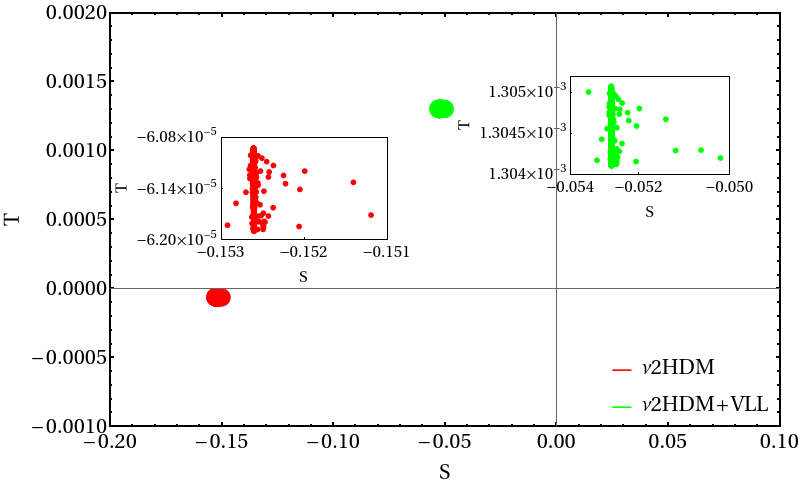}
    \caption{Scatter plot of $S$ and $T$ parameters for $\nu \rm 2HDM$ (red points) and $\nu \rm 2HDM$ + Vectorlike Leptons (green points) models.}
    \label{fig:STU}
\end{figure}
\end{itemize}

\subsection{The Vectorlike Lepton DM candidate}
To resolve the tension with the EWPT data, we add two vectorlike leptons (VLL) that couple with the Higgs boson doublets, specifically one vectorlike doublet $N$ and one vectorlike singlet $\chi$. 
If we assume both of the vectorlike fermions to be odd under the dark $Z_2^{\rm DM}$ symmetry (see Table \ref{table:1b}), then they can be stabilized as potential dark matter candidates, along with improving the oblique parameter limits. This charge assignment restricts the vectorlike leptons to mix with the SM leptons, which would have been the case if both (or even one) of them were even under $Z_2^{\rm DM}$. Therefore, in the subsequent analysis all the particles / fields displayed in both Table \ref{table:1} and Table \ref{table:1b} are considered.   

\begin{table}[h!]
\centering
\begin{tabular}{ |c|c|c|c|c|c| } 
\hline
Particle Name & $SU(2)_L$ Charges &  $U(1)_Y$ Charges & $Z_2$  Charges & $Z_2^{\rm DM}$ Charges\\
\hline
\multicolumn{5}{| c |}{\bf Fermionic Fields}\\
\hline
 N & 2 & -1  &  1  & -1\\
\hline 
$\chi$ &  1 & 0 & 1  & -1 \\ 
\hline
\end{tabular}
\caption{The Vectorlike lepton fields and their charge assignments. }
\label{table:1b}
\end{table}

The Lagrangian added to the neutrinophilic 2HDM is,
\begin{equation}
  L_{VLL} = m_N \bar{N} N + m_{\chi} \bar{\chi} \chi + y_N \bar{N} \tilde{\Phi_1} \chi + h.c., 
\end{equation} 
where the VLL doublet $N$ with hypercharge $Y = -1$ can be written as $N^T = (N_{0} \ \ N_{-})$ 
and the Yukawa terms initiate the mixing between $N_0$ and $\chi$ with hypercharge $0$, 
to finally produce two potential dark matter candidates $\chi_1$ and $\chi_2$. 
The Lagrangian in terms of the charged and neutral vectorlike leptons $N$ and $\chi$ is,    
\begin{eqnarray}
L_{VLL}^{mass} = m_N\bar{N_-}N_+ + (\bar{N_0} \ \ \bar{\chi}) \begin{pmatrix}
m_N & \frac{y_Nv_1}{\sqrt{2}}\\
\frac{y_Nv_1}{\sqrt{2}} & m_\chi
\end{pmatrix} 
\begin{pmatrix}
N_0\\
\chi
\end{pmatrix}.
\end{eqnarray}


This mass matrix involving the neutral fields can be further diagonalized to get the mass eigenstate ($\chi_1,\chi_2$). Assuming $N_0 = c_{\theta} \chi_2 - s_{\theta} \chi_1$, 
and $\chi = s_{\theta} \chi_2 + c_{\theta} \chi_1$ with $\theta$ being the mixing angle of 
these fermionic states, the masses of the physical eigenstates $\chi_1$ and $\chi_2$ are, 
\begin{eqnarray}
    m_{\chi_1}\;=\;m_{\chi}c_{\theta}^2+m_Ns_{\theta}^2-\frac{y_N~v_1}{\sqrt{2}}\sin2\theta \label{eq:2.38} \\ 
    m_{\chi_2}\;=\;m_Nc_{\theta}^2+m_{\chi}s_{\theta}^2+\frac{y_N~v_1}{\sqrt{2}}\sin2\theta \nonumber 
\end{eqnarray}
with $m_{\chi_1} < m_{\chi_2}$, and the mixing angle can be written as, 
\begin{eqnarray}
    \tan 2\theta \;=\; \;\frac{\sqrt{2}~(y_N~v_1)}{(m_N-m_{\chi})}.\label{eq:2.39}
\end{eqnarray} 
The model provides three free parameters: $m_N$, $m_{\chi}$ and $y_N$. Therefore, before we proceed to calculate the relic abundances involving these new particles, it is important to study the collider bounds available to these particles.

\begin{table}[htb!]
    \begin{minipage}{.5\linewidth}
      \centering
        \begin{tabular}{| c | c |} 
\hline
Coupling & Mathematical expression \\ [1ex]
\hline \hline
$\lambda_{\chi_1\chi_1 h}$ & $\frac{y_N}{2\sqrt{2}} s_\alpha\sin2\theta$  \\ [1ex]
\hline
$\lambda_{\chi_1\chi_1 H}$ &  $-\frac{y_N}{2\sqrt{2}} c_\alpha\sin2\theta$  \\ [1ex]
\hline
$\lambda_{\chi_2\chi_2 h}$ & $ -\frac{y_N}{2\sqrt{2}} s_\alpha\sin2\theta$  \\ [1ex]
\hline
$\lambda_{\chi_2\chi_2 H}$ & $\frac{y_N}{2\sqrt{2}} c_\alpha\sin2\theta$  \\ [1ex]
\hline
$\lambda_{\chi_1\chi_2 h}$ & $-\frac{y_N}{\sqrt{2}} c_\alpha\cos2\theta$ \\ [1ex]
\hline
$\lambda_{\chi_1\chi_2 H}$ & $ \frac{y_N}{\sqrt{2}} s_\alpha\cos2\theta$  \\ [1ex]
\hline
 \end{tabular}
    \end{minipage}%
    \begin{minipage}{.5\linewidth}
      \centering
       \begin{tabular}{|c|c|}
    \hline
    Coupling  & Mathematical expression \\ [1ex]
\hline \hline
$\lambda_{\chi_1 N_-W^+}$ & $ -\frac{e~\sin\theta}{\sqrt{2}~sin\theta_{w}}$  \\ [1ex]
\hline
$\lambda_{\chi_2 N_-W^+}$ & $ \frac{e~\cos\theta}{\sqrt{2}~sin\theta_{w}}$\\ [1ex]
\hline
$\lambda_{\chi_1 N_+W^-}$ & $ -\frac{e~\sin\theta}{\sqrt{2}~sin\theta_{w}}$ \\ [1ex]
\hline
$\lambda_{\chi_2 N_+W^-}$ & $ \frac{e~\cos\theta}{\sqrt{2}~sin\theta_{w}}$  \\ [1ex]
\hline
$\lambda_{\chi_1\chi_1 Z}$ & $ \frac{e~\sin^2{\theta}}{2\sin\theta_{w}~cos\theta_{w}}$ \\ [1ex]
\hline
$\lambda_{\chi_1\chi_2 Z}$ & $ \frac{e~\sin\theta~\cos\theta}{\sin\theta_{w}~\cos\theta_{w}}$ \\ [1ex]
        \hline
    \end{tabular}
    \end{minipage}
\caption{Coupling constants involving the SM and vectorlike DM particles.}
\label{tab:2}
\end{table}



{
{\underline{Bounds on Dark Matter particles} :} \\
We set $m_N$ = 3 TeV to be safe from the collider limits, which means that the heavier $m_{\chi_2}$ also becomes very heavy. Therefore, $m_{\chi_1}$ plays the role of a potential DM candidate. In fact, it is supposed to contribute as the second component of DM in our case study. The Higgs portal coupling $y_N$ contributes to the invisible decay widths of Higgs and $Z$ bosons. In Table \ref{tab:2}, we list the couplings relevant to our study, mainly those involving the two fermionic DM particles and the SM particles. After performing a scan in the $y_{N} - m_{\chi}$ plane, we find that a strong bound exists in the region with $y_{N} \ge$ 0.5 from the measurements of the invisible decay width. Therefore, we fix $y_N=$0.4 and vary $m_{\chi}$ in the range of $10 ~{\rm GeV}$ to $1~{\rm TeV}$, to obtain the physical masses and parameter-dependent coupling constants of the fermionic DM candidate. The scalar DM acquires a mass in the MeV range and therefore does not receive any strong bound from the colliders except its contribution to the Higgs invisible width. \\
We observe that direct detection of WIMP DM \cite{XENON:2018voc,PandaX-II:2017hlx,LUX:2017ree} imposes a stringent constraint on the mixing angle of the VLL DM singlet and doublet (Eq. \ref{eq:2.39}) $\sin \theta$ to be less than 0.05 \cite{Bhattacharyya:2015nca}. For the aforementioned choice of the Yukawa $y_N$, we find the mixing angle to be in the range $0.023 < \sin \theta < 0.034$ with $m_{\chi}$ in the $(10 - 1000)$ GeV mass range. Hence, the parameter space we have chosen is allowed by the direct-detection limits. Finally, we recalculate the S and T parameters for the extended model, $\nu 2HDM$ + vectorlike leptons, following the methodology of \cite{Ellis:2014dza}. The additional contribution from the VLLs helps to satisfy the bounds, as shown by the green points in Figure \ref{fig:STU}.}

Before we proceed to calculate dark matter relic density in the context of a 2-component dark matter framework, in Table \ref{tab:bp1} we display the details of the benchmark point (in terms of the masses and couplings) used for
the remaining analysis.  
\begin{table}[htb!]
\centering
\begin{tabular}{|c|c|c|c|c|c|c|c|}\hline
\multicolumn{8}{|c|}{Benchmark point} \\
\hline
$m_H$ & $m_h$ & $\alpha$ & $\tan\beta$  & $\kappa_1$ & $\kappa_2$ & $y_\nu$ & $m_{N_R}$ \\ 
\hline
20 MeV & 125 GeV & 89.999 & $8.13 \times 10^{-6}$ & 0.01 & 0.002 & $4.0\times 10^{-4}$ & 10 MeV  \\ [2ex] \hline
\end{tabular}
\vskip 0.5cm 
\begin{tabular}{|c|c|c|c|c|c|c|c|c|}\hline
\multicolumn{9}{|c|}{A few selected couplings} \\
\hline
	$\makecell{\lambda_{hHH}\\ \rm (GeV)}$ & $\makecell{\lambda_{hhH}\\ \rm (GeV)}$ & $\makecell{\lambda_{hhh}\\ \rm (GeV)}$ & $\makecell{\lambda_{\phi_3\phi_{3}h} \\ \rm (GeV)}$ & $\makecell{\lambda_{\phi_3\phi_{3}H}\\ \rm (GeV)}$  & 
    $\makecell{\lambda_{HW^- H^+}\\ \rm (GeV)}$  & 
    $\makecell{\lambda_{tbH^+}\\ \rm (GeV)}$  & 
    $\lambda_{\nu_L N_{R}h}$ & $\lambda_{\nu_L N_{R}H}$  \\ [2ex] \hline 
-1.04 & $1.36\times 10^{-6}$ & 63.51 & 2.47 & $2.26\times 10^{-5}$ & -0.326 & $-7.87 \times 10^{-6}$ & $-2.29 \times 10^{-9}$ & $2.82 \times 10^{-4}$ \\ [2ex] \hline
\end{tabular}
	\caption{{The benchmark scenario involved in our study. }} 
\label{tab:bp1}
\end{table}
\section{Dark Matter: Relic density Aspects} 
\label{sec:bz}

The freeze-out mechanism is an important artifact of thermal DM physics. In the early universe, most of its constituents were in thermal equilibrium with each other. However, as time progresses and the Universe cools down, the DM cannot interact enough to stay in thermal equilibrium with the SM. This instance of the decoupling of weakly interacting massive particles (WIMPs) determines the amount of dark matter present at the current Universe, which for a vanilla DM case is termed the freeze-out. To incorporate the freeze-out phenomenon in the thermal history of the Universe, we study the comparison of the interaction rate of the particle with the expansion rate of the Universe, denoted by the Hubble constant H. As the interaction rate $\Gamma$ of the particles gradually decreases and becomes smaller than the Hubble constant, this evolution and eventual decoupling from the SM plasma are incorporated through a Boltzmann equation~\cite{Kolb:1990vq}.

In the model set up we motivated the presence of two possible dark matter candidates, one fermionic DM which is relatively heavy (mass around $\sim 100~$GeV) while the other one a light scalar with mass around $10~$MeV. The fermionic DM $\chi_1$ is a vectorlike fermion, while the scalar DM $\phi_3$ is a singlet scalar. Both candidates for DM are analyzed using a thermal DM treatment, where the abundance of DM is set up through a freeze-out mechanism. Here, we first discuss the Boltzmann equation for the single-component DM candidates: both the scalar DM and fermion DM candidates separately, assuming that one of them is present in the model at one instant. After that, we explore the case where both the fermionic and the scalar DM candidates are present together and interact with each other. Their thermal evolution is connected through a set of coupled Boltzmann equations. In each of these scenarios, the Boltzmann equations are solved to find the freeze-out temperature (defined through $x_f \equiv m/T_f$) and using this temperature the corresponding relic abundances are computed.

\subsection{Single-component DM} \label{sec:signle-DM}
For each of the DM components, the relic density is determined by the Boltzmann equation which is driven by the corresponding DM annihilation cross section. Here we assume that there is no interaction between the two potential dark matter candidates $\phi_3$ and $\chi_1$, as only one of them can be introduced, building a consistent theory. Both have their own role in the phenomenology of the neutrinophilic model. We discuss the scalar DM first, then we study the fermionic DM, and we discuss the results together. 
\begin{itemize}
%
\item {\bf Scalar DM $\phi_3$} \\
{The scalar DM here is a singlet scalar $\phi_3$ introduced in the neutrinophilic two Higgs doublet model. For single component DM, the Boltzmann equation is written as: 
\begin{equation}
\frac{dn_{\phi_3}}{dt} + 3 H n_{\phi_3} = -  \langle \sigma_{\phi_3} v \rangle (n_{\phi_3}^2 - n_{\phi_3,\rm eq}^2) \label{eq:4.1}
\end{equation}
		Here $ \sigma_{\phi_3}$  is the annihilation cross section of the scalar dark matter candidate, when $\phi_3$ annihilates to the visible sector (as shown in Fig.~\ref{fig:scalarDM-feynman-diag}), expressed as,
\begin{eqnarray}
   \sigma_{\phi_3}(s) &=& \sigma(s)_{\phi_3\phi_3\to f \Bar{f}}+ \sigma(s)_{\phi_3\phi_3\to N_R \nu_L}+ \sigma(s)_{\phi_3\phi_3\to HH}+ \sigma(s)_{\phi_3\phi_3\to hh}+ \sigma(s)_{\phi_3\phi_3\to hH} , \nonumber  
\end{eqnarray}
\begin{figure}[h!]
    \centering
    \includegraphics[scale=0.7]{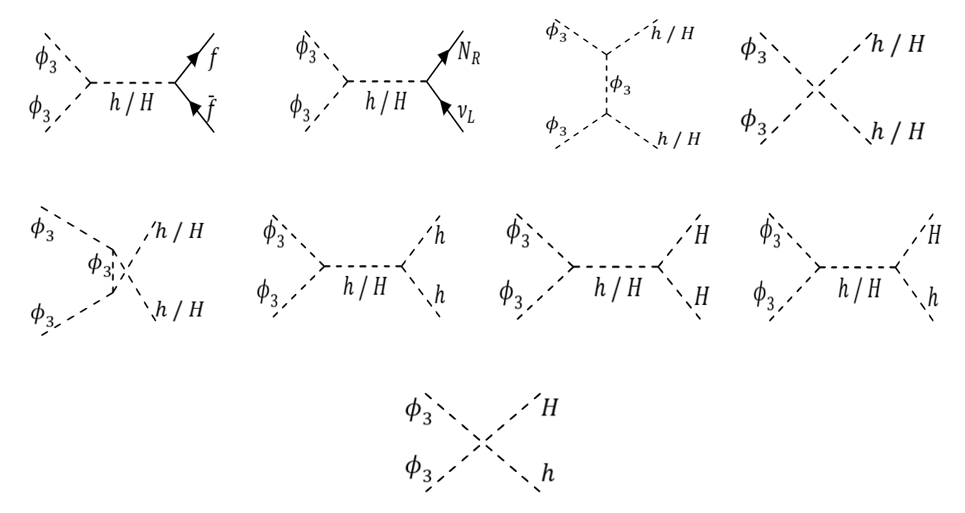}
    \caption{Feynman diagrams for annihilation processes of scalar DM $\phi_3$.}
    \label{fig:scalarDM-feynman-diag}
\end{figure}
		where $\langle \sigma_{\phi_3} v \rangle$ is the thermal averaged cross section. The mathematical expression of the cross section for each process is given in the Appendix \ref{sec:A.1}. Thermal evolution of the DM, together with the DM relic density computation requires the thermal-averaged annihilation cross section as an input. 

For a non-relativistic regime, the whole process of calculating a thermal averaged cross section boils down to $\langle\sigma v\rangle\equiv \sigma v$~\cite{Bauer:2017qwy} through s-wave scattering analysis~\cite{Gondolo:1990dk}. Here, $v$ is the relative velocity between two DM particles, given through $ s = m_{\phi_3}^2 v^2 + 4m_{\phi_3}^2$. Expanding the term $\sigma v$ in Taylor series for $x \gg 1$, i.e. in the non-relativistic limit, we get 
\begin{align}
    \langle \sigma v \rangle~=~\sigma_0~+~\frac{\sigma_1}{x} ,
\label{eq:4.2}
\end{align}
where, parameter~$x = m_{\phi_3}/T$, T being the temperature of the thermal bath. In the case of light dark matter as is the case for the scalar DM here, the DM has the potential chances to be boosted. At the same time, one can also have a light DM which is not boosted, for which the thermal averaging mechanism is outlined above. We later show that, in the context of the model described above, $\phi_3$ can actually be produced through the $\chi_1~\chi_1 \rightarrow \phi_3~\phi_3$ process, which will lead to boosted DM due to kinematic factors. The effect of the boost is explored separately in the next section Sec.~\ref{sec:detection-boost}.
Hence, considering the fact that $\phi_3$ here may or may not get enough velocity to become relativistic, we decide to go ahead with the most general thermal averaged cross section computation method. For the scalar DM $\phi_3$, the thermal averaged cross section is given by,
\begin{eqnarray}
{\langle \sigma v \rangle}_{\phi_3\phi_3 \to SM} = \frac{x}{8m_{\phi_3}^5 K_2^2\Bigl(x\Bigl)}~\int_{4m_{\phi_3}^2}^{\infty}~\sigma(s)_{\phi_3} \times (s-4m_{\phi_3}^2) \sqrt{s} K_1\Bigl(\frac{x\sqrt{s}}{m_{\phi_3}}\Bigl) ds,
\label{eq:4.3}
\end{eqnarray}
where $K_n$ are modified Bessel functions of order n. In Eq.~\ref{eq:4.1}, we can scale the number density with respect to the total entropy of the Universe, $s$, to work with a quantity called the comoving density $Y_i = n_i/s$ that evolves with respect to a parameter $x = m/T$, T being the temperature of the Universe. With this scaling, the Boltzmann equations are modified as:
\begin{eqnarray}
\frac{dY_{\phi_3}}{dx}  &=& -\frac{1}{x^2} \frac{s(m_{\phi_3})}{H(m_{\phi_3})}  \langle\sigma v\rangle_{\phi_3\phi_3\to SM} (Y_{\phi_3}^2 - Y_{\phi_3,\rm eq}^2) 
  \label{eq:4.4}
\end{eqnarray}

where, we choose $Y_{\phi_3,\rm eq}=0.145~\frac{g_i}{g*}~x^{3/2}~e^{-x}$ with the internal degrees of freedom for scalar DM $g_i=1$, and the effective number of degrees of freedom $g*=10$ is considered for the MeV scale freeze-out temperature \cite{Kolb:1990vq}. Notice that the lower limit of the integration over the CM energy is $4m_{\phi_3}^2$ for the thermal averaged cross section. Note that a particular DM particle cannot annihilate to a particle that has a mass greater than the DM mass. 
The reason being the fact that the cross section of the DM annihilation contains a factor $\sqrt{s-4m_i^2}$, which ceases to be real when $m_{\rm DM} \le m_i$, $m_i$ being the mass of $i$-th particle, that the DM pairs can annihilate to. Therefore, based on the constraint, only kinematically available annihilation channels will be open for light-scalar DM $\phi_3$ annihilation.
}

\item {\bf Fermion DM $\chi_1$} \\
{For the vectorlike fermion component of the DM, the equation takes the form,
\begin{equation}
\frac{dn_{\chi_1}}{dt} + 3 H n_{\chi_1} = - \frac{1}{2} \langle \sigma_{\chi_1} v \rangle (n_{\chi_1}^2 - n_{\chi_1,\rm eq}^2) \label{eq:4.5}
\end{equation}
Here $\sigma_{\chi_1}$ is the annihilation cross section of the fermionic dark matter with $\chi_1$ annihilating to the visible sector as:
\begin{eqnarray}
   \sigma_{\chi_1}(s) &=& \sigma(s)_{\chi_1\chi_1\to f\bar{f}}+ \sigma(s)_{\chi_1\chi_1\to N_R\nu_L}+ \sigma(s)_{\chi_1\chi_1\to hh}+ \sigma(s)_{\chi_1\chi_1\to ZZ} \nonumber \\ &+& \sigma(s)_{\chi_1\chi_1\to W^+ W^-} \label{eq:4.6}
\end{eqnarray}
and $\langle \sigma_{\chi_1} v \rangle$ are the thermal averaged cross section. The Feynman diagrams are presented in Fig.~\ref{fig:fermionDM-feynman-diagrams}, and the cross section of each annihilation process is described in Appendix~\ref{sec:A.2}.
\begin{figure}[h!]
    \centering
    \includegraphics[scale=0.7]{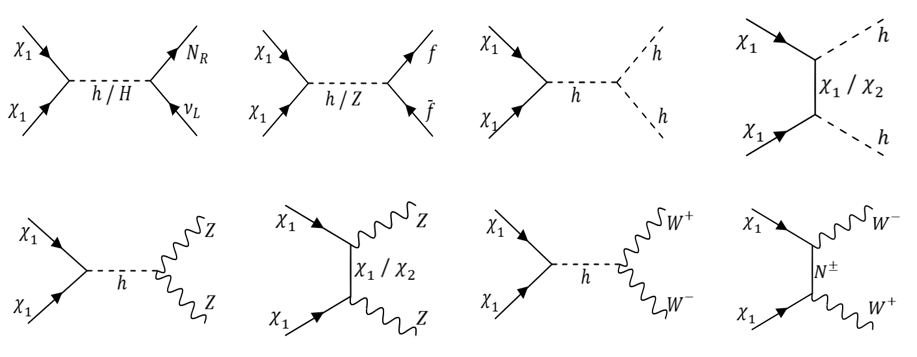}
    \caption{Feynman diagrams for annihilation processes of fermionic DM $\chi_1$.}
    \label{fig:fermionDM-feynman-diagrams}
\end{figure}

 Like in the previous case, the thermal averaged cross section of the fermionic DM, $\chi_1$ is given by,
\begin{eqnarray}
{\langle\sigma v \rangle}_{\chi_1\chi_1\to SM} = \frac{x}{8m_{\phi_3}^5K_2^2\Bigl(x\Bigl)}~\int_{4m_{\chi_1}^2}^{\infty}~\sigma(s)_{\chi_1} \times (s-4m_{\chi_1}^2) \sqrt{s} K_1\Bigl(\frac{x\sqrt{s}}{m_{\chi_1}}\Bigl) ds   \label{eq:4.7}
\end{eqnarray}
Scaling the number density with respect to the total entropy of the Universe, $s$, be defining a comoving density 
$Y_{\chi_1} = n_{\chi_1}/s$ and it's evolution with respect to a parameter $x = m/T$, the modified Boltzmann equation for fermionic DM 
is modified to,

\begin{eqnarray}
  \frac{dY_{\chi_1}}{dx} &=&  -\frac{1}{2} \frac{1}{x^2}\frac{s(m_{\chi_1})}{H(m_{\chi_1})}\langle\sigma v\rangle_{\chi_1\chi_1\to SM} (Y_{\chi_1}^2 - Y_{\chi_1,\rm eq}^2)  
  \label{eq:4.8}
\end{eqnarray}
where, we set $Y_{\chi_1,\rm eq}=0.145~\frac{g_i}{g*}~x^{3/2}~e^{-x}$, with the internal degrees of freedom for fermionic DM $g_i=2$, and the effective number of degrees of freedom $g*=80$ is considered for the GeV scale freeze-out temperature \cite{Kolb:1990vq}.
}
\\

\item {\bf Numerical Results} \\
	{The solution of the Boltzmann Equation for scalar DM (Eq.~\ref{eq:4.4}) and for fermionic DM (Eq.~\ref{eq:4.8}) candidates separately gives the Fig.~\ref{fig:scalar-DM} and Fig.~\ref{fig:fermion-DM} respectively. We show the freeze-out plot for one benchmark point with a fixed mass of both the DM candidates, $\phi_3$ and $\chi_1$. Solving Eq.~\ref{eq:4.9} and following Eq.~\ref{eq:4.11}, we get the freeze-out temperature, DM density at the freeze-out and consequent relic abundance for the scalar DM $\phi_3$ and fermionic DM $\chi_1$ which are plotted in the right panel of Fig.~\ref{fig:scalar-DM} and \ref{fig:fermion-DM} respectively. The relic abundance, can be expressed as~\cite{Drees:2021rsg} 
\begin{eqnarray}
    \Omega h^2= \frac{2.14\times10^9~\rm{GeV^{-1}}}{\sqrt{g_\ast}~M_{pl}}\frac{1}{J(x_f)} \label{eq:4.9}
\end{eqnarray}
where $M_{\rm Pl}=1.22\times 10^{19}~\rm{GeV}$ is the Plank Mass and $J(x_f)$ is given by 
\begin{eqnarray}
    J(x_f)=\int_{x_f}^{\infty}~\frac{\langle\sigma v\rangle(x)}{x^2}~dx \label{eq:4.10}
\end{eqnarray}
where $x_f=m/{T_f}$ with $T_f$ be the freeze-out temperature, obtained solving the Boltzmann equation governing DM thermal evolution. 

\begin{figure}[h!]
\centering
\begin{subfigure}{0.5\textwidth}
  \centering
  \includegraphics[width=0.9\linewidth]{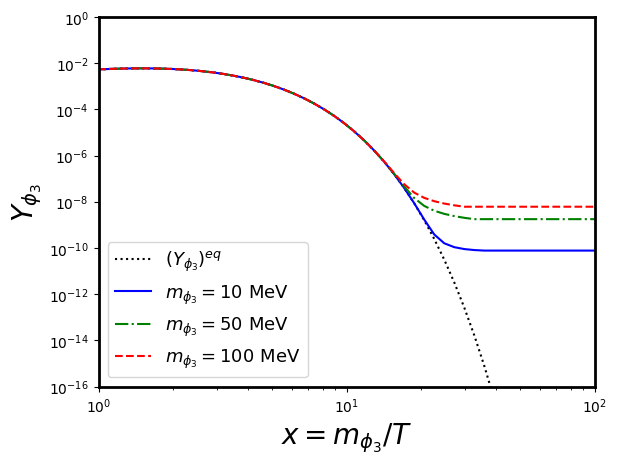}
\end{subfigure}%
\begin{subfigure}{.5\textwidth}
  \centering
  \includegraphics[width=0.9\linewidth]{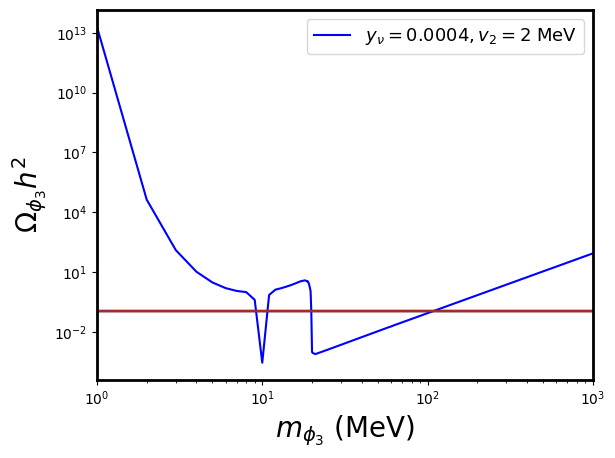}%
\end{subfigure}
\caption{{Left: Illustration of the freeze-out phenomenon for scalar dark matter $\phi_3$, where the dark matter annihilation determines its thermal evolution through single component Boltzmann equation. Right: Scalar dark matter relic density ($\Omega_{\phi_3} h^2$) variation with mass $m_{\phi_3}$. The red band shows the relic density observed by the Planck experiment.}}
\label{fig:scalar-DM}
\end{figure}

In Fig.~\ref{fig:scalar-DM} we present the phenomenology of a scalar DM. In these plots, a single component Boltzmann Equation is solved for individual scalar DM $\phi_3$, to obtain the timeline and nature of freeze-out phenomenon for $\phi_3$. The left panel of Fig.~\ref{fig:scalar-DM} is drawn for different values of $m_{\phi_3}$, with other model parameters being fixed to particular benchmark values. We obtain the freeze-out phenomenon around $x \sim 15-20$, with the largest x obtained for a scalar dark matter mass $m_{\phi_3} \approx 10 ~$MeV. This is a sign of a relatively late freeze-out which happens due to enhanced annihilation, dominantly aided due to the DM mass being around the H-resonant region.
%


{The right panel of Fig.~\ref{fig:scalar-DM} illustrates the variation of the scalar DM relic density as a function of the DM mass $m_{\phi_3}$. Due to the small values of the Yukawa coupling ($y_{\nu}$) and the vev ($v_2$), the dominant $H$ mediated annihilation channel $\phi_3 \phi_3 \to N_R \nu_L$, produces suppressed cross-section; leads to over-abundant relic density even at the vicinity of the resonance $m_{\phi_3} \simeq m_H/2 = 10$ MeV. However, at the exact resonance, the annihilation cross-section is significantly enhanced, leading to a sharp drop in the relic density, aligning with the experimentally observed relic abundance.\\
Beyond the resonance, the relic density remains large until a second sharp drop occurs around $m_{\phi_3} \sim m_H = 20$ MeV. This drop marks the opening of the t-channel annihilation process $\phi_3 \phi_3 \to H H$, which plays a crucial role in reducing the relic density. This annihilation channel provides an efficient depletion mechanism and helps to evade over-abundance constraints, thereby salvaging the viability of the model. For $m_{\phi_3}>m_H$, no additional significant annihilation channels contribute, and the relic density gradually increases with increasing DM mass $m_{\phi_3}$ or the increasing center-of-mass energy $s=4 m_{\phi_3}^2$. The red band shows the 3$\sigma$ limit obtained from the experimentally measured of the relic density by the Planck collaboration \cite{Planck:2015fie}.
}


Following a non-relativistic approach for the heavier DM, the $\langle \sigma v \rangle$ is not a function of $x$. 
Therefore, the relic density expression for the fermionic DM becomes,
\begin{eqnarray}
\Omega h^2= \frac{2.14\times 10^9~x_f}{\sqrt{g_\ast}M_{Pl}~\langle \sigma v\rangle} 
\label{eq:4.11}
\end{eqnarray}
Using this, Eq.~\ref{eq:4.9} and Eq.~\ref{eq:4.11} are followed here to plot freeze-out diagrams and to find the relic density for the fermionic DM $\chi_1$ in Fig.~\ref{fig:fermion-DM}.
\begin{figure}[h!]
\centering
\begin{subfigure}{0.5\textwidth}
  \centering
  \includegraphics[width=0.93\linewidth]{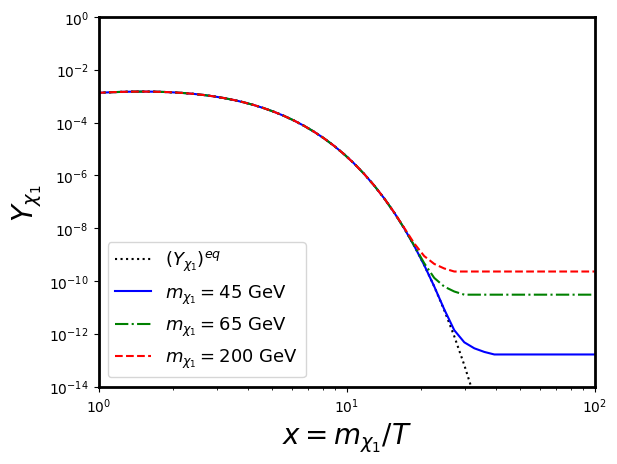}
  \label{fig:a}
\end{subfigure}%
\begin{subfigure}{.5\textwidth}
  \centering
  \includegraphics[width=0.93\linewidth]{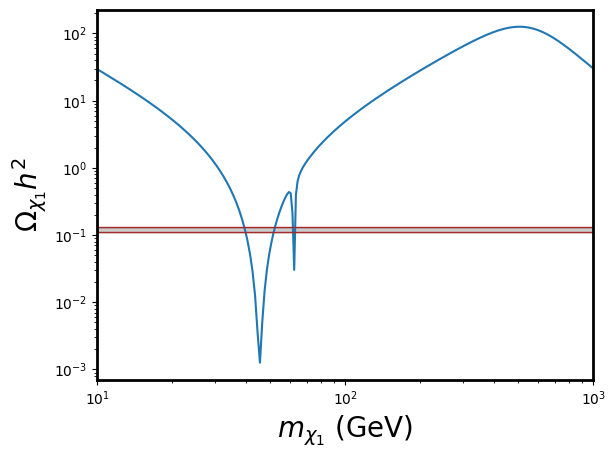}
  \label{fig:b}
\end{subfigure}
\caption{{Left: Illustration of the freeze-out phenomenon through solving the Boltzmann Equation for the single component fermionic DM $\chi_1$, with freeze-out phenomenon at different masses of $\chi_1$. 
Right: DM relic density with the variation of fermionic DM mass.
The red band shows the relic density observed by the Planck experiment.}}
\label{fig:fermion-DM}
\end{figure}

\begin{table}[h!] 
\centering
\begin{tabular}{|c|c|c|c|} \hline
\multicolumn{2}{|c|}{Scalar DM} & \multicolumn{2}{|c|}{Fermion DM}\\ \hline
Mass & Relic density &  Mass & Relic density \\ [2ex] \hline 
10 MeV & $1.92\times 10^{-4}$ & 45 GeV & $2.07 \times 10^{-3}$ \\ [2ex] \hline 
50 MeV & $1.07\times 10^{-2}$ & 65 GeV & 0.67 \\ [2ex] \hline 
100 MeV & $8.59\times 10^{-2}$ & 200 GeV & 22.16 \\ [2ex] \hline 
\end{tabular}
\caption{{Individual relic densities of the scalar and fermionic dark matter in single component dark matter scenario, corresponding to different dark matter masses in both of the cases.}}
\label{tab:relic-density}
\end{table}

The relic density for the fermionic DM is shown in Fig.~\ref{fig:fermion-DM} (right), where the two resonance drops in the relic density happen due to the contribution coming from the $Z$ and $h$ mediated $s$ channel diagrams respectively, with the $Z$ resonance drop being way more significant that the other. These two resonant drops observed respectively at $m_{\chi_1} \sim m_Z/2$ and $m_{\chi_1} \sim m_h/2$ are evident from Fig.~\ref{fig:fermion-DM} (right). Note, we have considered the Breit-Wigner resonance width of 2.5 GeV and 4 MeV for the $Z$ and $h$ bosons respectively. The relevant Feynman diagrams are shown in Fig.~\ref{fig:fermionDM-feynman-diagrams} and in Sec.~\ref{sec:A.2}, we present the expression for the cross section for each diagrams of the $\chi_1$ annihilation. 
A similar conclusion can be drawn from Fig.~\ref{fig:fermion-DM} where only a small region of fermionic DM, $\chi_1$ mass is allowed with an underabundant or exact relic density, touching the experimentally measured relic density band. The observations of Fig.~\ref{fig:scalar-DM} and Fig.~\ref{fig:fermion-DM} are highlighted in Table~\ref{tab:relic-density} displaying a few points for different masses of $\phi_3$ and $\chi_1$ and their corresponding relic densities, respectively. 

}

\end{itemize}

\subsection{Two-Component Dark Matter: Coupled Boltzmann Equation} \label{sec:CBE}
\label{coupledBE}
\begin{figure}[h!]
    \centering
    \includegraphics[scale=0.4]{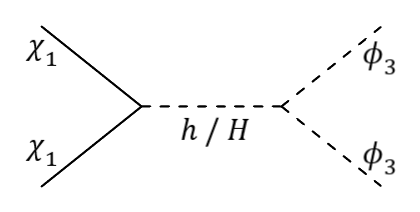}
    \caption{Higgs boson (h/H) mediated $\chi_1\chi_1 \to \phi_3\phi_3$ annihilation diagram. }
    \label{fig:FSanni}
\end{figure}

In the two-component dark matter model, which is under study, the fermionic dark matter is found to be heavier and it can annihilate to lighter scalar dark matter. 
Their relic density together can be obtained only when we consider their number densities to evolve through a set of coupled Boltzmann equations (CBE). 
The Boltzmann equations are expressed as:
\begin{eqnarray}
\frac{dn_{\chi_1}}{dt} + 3 H n_{\chi_1} &=& - \frac{1}{2} \langle 
\sigma_{\chi_1\chi_1 \to \phi_3\phi_3} v \rangle (n_{\chi_1}^2 - n_{\phi_3}^2 
\frac{n_{\chi_1,\rm eq}^2}{n_{\phi_3,\rm eq}^2}) - \frac{1}{2} \langle \sigma_{\chi_1} v \rangle (n_{\chi_1}^2 - n_{\chi_1,\rm eq}^2)  \nonumber \\
\frac{dn_{\phi_3}}{dt} + 3 H n_{\phi_3} &=& -  \langle \sigma_{\phi_3} v \rangle (n_{\phi_3}^2 - n_{\phi_3,\rm eq}^2)  -  \langle \sigma_{\phi_3\phi_3 \to \chi_1\chi_1} v \rangle (n_{\phi_3}^2 - n_{\chi_1}^2 \frac{n_{\phi_3,\rm eq}^2}{n_{\chi_1,\rm eq}^2}) 
\label{eq:cbe1}
\end{eqnarray}
Here, $\sigma_{\chi_1\chi_1 \to \phi_3\phi_3}$ provides the cross section of the fermionic DM annihilating to the scalar DM, whereas 
$\sigma_{\phi_3\phi_3 \to \chi_1\chi_1}$ provides the same for the reverse process\footnote{The expression of the cross sections are given in Appendix.~\ref{eq:A.6}}. 
 The $\chi_1\chi_1\to \phi_3\phi_3$ annihilation, which results due to the interaction between the two types of DM candidates is represented by the Feynman diagram in Fig.~\ref{fig:FSanni}. 
Here the $H$ mediated process is highly suppressed due to very weak interaction strength compared to the interactions involving the SM Higgs at the alignment limit. Two important points to note here. First, there could be other processes involving SM particles which can produce the $\phi_3$ particles, for example $f\Bar{f} \to \phi_3 \phi_3$, $W^+W^-\to \phi_3\phi_3$, $ZZ \to \phi_3 \phi_3$, $hh\to \phi_3 \phi_3$ etc., however we have checked that all these processes are highly suppressed compared to the most dominant process $\chi_1 \chi_1 \to \phi_3 \phi_3$. It is implied then that almost all of these $\phi_3$ particles, produced through the $\chi_1$ annihilation process, are sufficiently boosted, and therefore, the coupled Boltzman Equations don't require any modification. Secondly, it is also true that there could be other modes of annihilation process for the $\chi_1$ other than $\chi_1 \chi_1 \to \phi_3 \phi_3$. Again we checked that the 
$\chi_1 \chi_1 \to \phi_3 \phi_3$ and $\chi_1 \chi_1 \to f \bar{f}$ are the two most dominant modes of $\chi$ annihilation, however $\chi_1 \chi_1 \to f \bar{f}$ process is one-order suppressed compared to the former.

 Scaling Eq.~\ref{eq:cbe1} with the comoving density $Y_i=n_i/s$ and the parameter $x \equiv x_{\phi_3}= m_{\phi_3}/T$ we get,
\begin{eqnarray}
\frac{dY_{\chi_1}}{dx} &=& - \frac{1}{2}\frac{\lambda_{\chi \phi}}{x^2}
\left(Y_{\chi_1}^2 - Y_{\phi_3}^2 \frac{Y_{\chi_1,\rm eq}^2}{Y_{\phi_3,\rm eq}^2} \right) - \frac{1}{2} \frac{\lambda_{\chi}}{x^2}
\left(Y_{\chi_1}^2 - Y_{\chi_1,\rm eq}^2 \right)  \nonumber \\
\frac{dY_{\phi_3}}{dx}  &=& -\frac{\lambda_{\phi}}{x^2}
\left(Y_{\phi_3}^2 - Y_{\phi_3,\rm eq}^2 \right)  + \frac{\lambda_{\chi \phi}}{x^2}
\left(Y_{\chi_1}^2 - Y_{\phi_3}^2 \frac{Y_{\chi_1,\rm eq}^2}{Y_{\phi_3,\rm eq}^2} \right) \nonumber \\ \label{eq:cbe2}
\end{eqnarray} 
with 
$$ \lambda_{\chi \phi} = \frac{s(m_{\phi_3})}{H(m_{\phi_3})} \langle\sigma v\rangle_{\chi_1\chi_1 \to \phi_3\phi_3}, $$ 
$$ \lambda_{\chi} = \frac{s(m_{\phi_3})}{H(m_{\phi_3})}\langle\sigma_{\chi_1} v\rangle \hspace{0.5cm}
 \lambda_{\phi}  = \frac{s(m_{\phi_3})}{H(m_{\phi_3})}  \langle\sigma_{\phi_3} v\rangle $$

\begin{figure}[h!]
\centering
\begin{subfigure}{0.5\textwidth}
  \centering
  \includegraphics[width=0.9\linewidth]{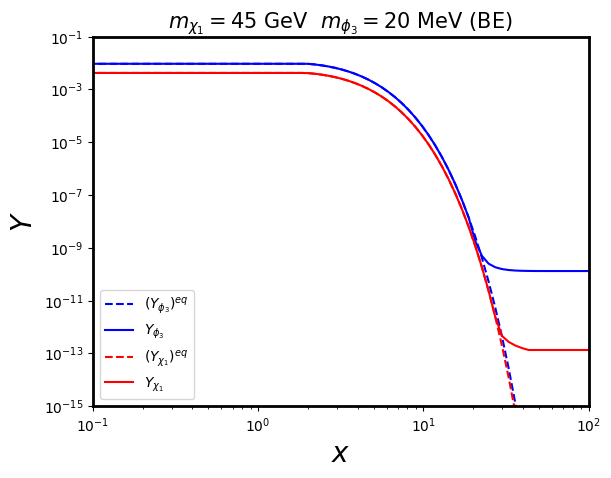}
  \label{fig:a}
\end{subfigure}%
\begin{subfigure}{.5\textwidth}
  \centering
  \includegraphics[width=0.9\linewidth]{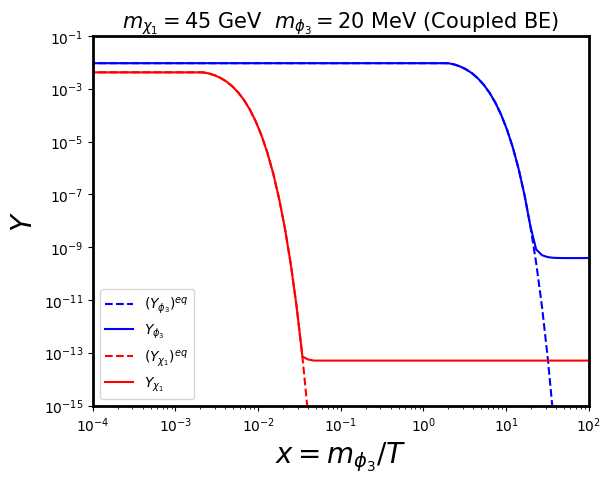}
  \label{fig:b}
\end{subfigure}
\caption{{Freeze-out plot for Coupled Boltzmann Equation (right panel) with  $m_{\chi_1}=45~{\rm GeV}$ $m_{\phi_3}=20~{\rm MeV}$ and its comparison with the case of individual DM candidates (left panel).}}
\label{fig:CaseI-65GeV}
\end{figure}

\begin{figure}[h!]
\centering
\begin{subfigure}{0.5\textwidth}
  \centering
  \includegraphics[width=0.95\linewidth]{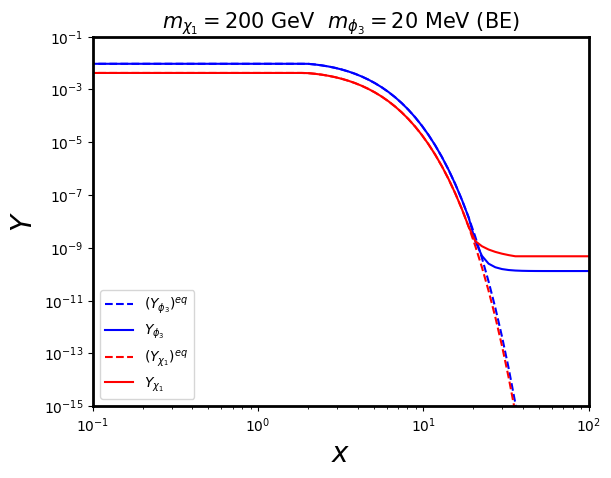}
  \label{fig:a}
\end{subfigure}%
\begin{subfigure}{0.5\textwidth}
  \centering
  \includegraphics[width=0.95\linewidth]{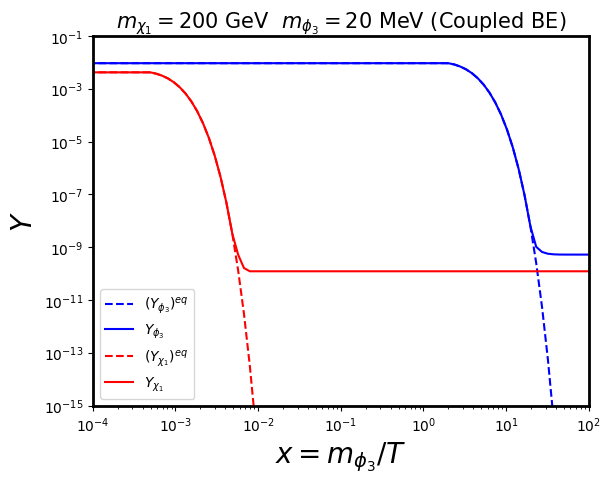}
  \label{fig:b}
\end{subfigure}
\caption{{Freeze-out plot for coupled Boltzmann equation (right panel) with $m_{\chi_1}=200~{\rm GeV}$ and $m_{\phi_3}=20~{\rm MeV}$, and its comparison with the case of individual DM candidates (left panel).
}}
\label{fig:CaseII-200GeV}
\end{figure}

{where $ s(m_{\phi_3})=\frac{2\pi^2}{45}g^*m_{\phi_3}^3$, $H(m_{\phi_3})=\frac{\pi}{90}\frac{\sqrt{g^*}}{M^{r}_{\rm Pl}} m_{\phi_3}^2$ with reduced plank mass $M^{r}_{\rm Pl} = 2.44 \times 10^{18}~{\rm GeV}$, and $\langle \sigma v\rangle$ is the thermal averaged annihilation cross section as discussed before. Here the equilibrium number density for the i-th particle is considered as \cite{Kolb:1990vq},
\begin{eqnarray} \nonumber
    Y_{i,eq} &=& 0.278 ~\frac{g_{eff}}{g*} \quad (x<3) \\ \nonumber
    Y_{i,eq} &= &0.145~\Big(\frac{g_i}{g*}\Big)\Big(\frac{m_i}{m_{\phi_3}}\Big)^{3/2}~x^{3/2}~e^{-\frac{m_i}{m_{\phi_3}}x} \quad  (x \ge 3)\nonumber 
\end{eqnarray}
where, $g_{eff}=g_i$ (for boson with $g_i=1$) and $g_{eff}=3g_i/4$ (for fermion with $g_i=2$). And the $g*=10$ for a MeV scale $T_{f}$ and $g*=80$ for a GeV scale $T_{f}$. 
}In the center of mass (CM) frame, energy can be written in terms of the DM mass and $x$ parameter, such that $s=4m_{\chi_1}^2+ 2 m_{\chi_1}^2/x$. The heavy fermionic dark matter obeys the non-relativistic limit as the parameter $x=m_{\chi_1}/T \gg 1$ and $s \approx 4~m_{\chi_1}^2$. Hence, we implement the approximation $\langle \sigma v\rangle = \sigma \   v$ for the non-relativistic regime while considering the fermionic DM.
Based on the earlier discussion, in Eq.~\ref{eq:4.2} we have taken $\langle \sigma_{\chi_1} v \rangle~=~\sigma_0 + \sigma_1/x~ \cong ~\sigma_0$, with $x$ being very very large 
$\sigma_0 \approx \sigma_{\chi_1}(s=4 m_{\chi_1}^2)$. 
Next we study two benchmark scenarios -- we compare the freeze-out of the individual dark matter candidates where they do not interact and the two-component dark matter scenario, where the interaction term is present. The first scenario is already analyzed in detail in the previous subsection where we discuss the single component DM scenarios. 
To study the two-component scenario, we have used the coupled Boltzmann equation (CBE), i.e., Eq.~\ref{eq:cbe2}, which provides the freeze-out plots of the fermionic DM $\chi_1$ and the scalar DM $\phi_3$ together.
The general feature of the single DM and the coupled two-component DM scenario is presented in the  Scenario-I (Figure~\ref{fig:CaseI-65GeV}) and  Scenario-II (Figure~\ref{fig:CaseII-200GeV}). 
The left diagrams for both the Scenarios depict the freeze-out for the individual fermionic DM $\chi_1$ and the scalar DM $\phi_3$, both of them as the solution of the corresponding single component Boltzmann Equation (BE).
\footnote {For the two DM coupled cases, two $x$ variables are possible: we choose $x = m_{\phi_3}/T$. 
We observed that, by this choice, the freeze-out happens for the scalar and fermionic DM at different $x$ values, the fermion DM freezes out at a very smaller $x \sim 10^{-2}$ values while the scalar DM freezes out at $x \sim 15$. This fermionic DM freeze out $x =m_{\phi_3}/T  \sim 10^{-2}$ can easily be translated to a x value $x =m_{\chi_1}/T  \sim 10$ which puts the freeze out timeline of both the DM candidates in a similar temperature regime. 
}

The results on the left show that a $\sim 45$~GeV fermionic DM can have an efficient resonant annihilation to have a relatively lower yield at the freeze-out, while at higher DM masses like $m_{\chi_1} \sim 200$ GeV it goes to relic values that is over-abundant. 
For the scalar DM, relic density at freeze-out will be relatively higher as the DM relic remains in the experimentally allowed range around 
$m_{\phi_3} \sim 20-100$ MeV, while it goes further down at exact resonant region, explored at $m_{\phi_3} \sim 10~$MeV.
We have shown the coupled two-component DM cases in Fig.~\ref{fig:CaseI-65GeV} and Fig.~\ref{fig:CaseII-200GeV}. In general we have found that:
\begin{itemize}

\item Compared to the single component DM case, $x_f$ for the fermion DM sees a minor increase: it is an artefact of increased annihilation leading to late freeze-out. 

\item The fermionic DM in the coupled sector has one additional mode to annihilate to which reduces the relic density, which is reflected in the lower DM yield. 

\item The dark matter yield for the fermionic DM decreases for both the scenarios leading to lower $Y_{\infty}$ ($Y_{\infty}=Y_{eq}(x_f)$), compared to the individual DM cases. 
The scalar DM yield remains the same albeit increases a bit in the coupled scenario. 
That happens as the scalar DM gets boosted being produced from the fermionic DM annihilation. 

\item This boost is not always sufficient enough to initiate new annihilation modes, which was hitherto unavailable due to kinematic conditions. 
Still, due to the boost effects the s-channel resonant annihilation becomes less efficient which increases the relic.

 \end{itemize}
\paragraph {\textbf{Scenario-I:} }
 In this case (shown in Fig.~\ref{fig:CaseI-65GeV}) we take the fermionic DM mass in its $Z$ mediated resonant annihilation region, where its individual relic density will be
under-abundant. DM yield obtained is even lower compared to the way it happens in the Higgs mediated resonance region. The scalar DM mass is taken as $m_{\phi_3} \sim 20$~MeV which is away from the resonant region of $m_H/2 \sim 10~$MeV but the relic density remains under-abundant. {This is due to the onset of t-channel $\phi_3\phi_3 \to HH$ annihilation channel at $m_{\phi_3} \simeq m_{H}$.} When we look at their individual dark matter phenomenology in Fig.~\ref{fig:CaseI-65GeV}(Left), the scalar DM freezes out at a smaller $x$ but with a larger $Y$ compared to the Fermionic DM.
The large $Y_{\phi_3}$ manages to provide the DM relic lower than the fermionic DM by one order of magnitude. In contrast, 
for the fermionic DM, $Y_{\chi_1}$ is smaller by one order of magnitude compared to $Y_{\phi_3}$, which results in a larger relic density compared to the scalar DM.
In the two-component dark matter case Fig.~\ref{fig:CaseI-65GeV}(Right), the fermionic dark matter yield $Y_{\chi_1}$, decreases further due to its extra annihilation to the scalar DM mode. 
Even if the scalar DM is boosted, it cannot gather enough energy to include any additional annihilation channels in this scenario. 
But its s-channel resonance contribution gets diluted due to a modified resonance condition in presence of a boost. As a result, its annihilation decreases and yield $Y_{\phi_3}$ increases further. 
This is reflected in the increased gap between the two freeze-out lines, corresponding to the fermionic and scalar DM cases in the coupled DM scenario.
Hence, the fermionic contribution to the DM relic comes down in the coupled case compared to the single component fermionic DM case for the same mass.
We saw in Fig.~\ref{fig:scalar-DM} and Fig.~\ref{fig:fermion-DM} that the individual DM relic densities were far apart, but now in 
Fig.~\ref{fig:CaseI-65GeV}(Right) we show that the scalar DM relic density increases and it is only one order of magnitude 
higher than the fermionic one.
Overall, the total DM density goes down for the coupled DM case compared to the same scenario with two individual DM candidates, small enough to be 
under-abundant but relatively closer to the exact relic.\\
\paragraph {\textbf{Scenario-II:}}
In this case (shown in Fig.~\ref{fig:CaseII-200GeV}) we consider the fermionic DM mass 
in a region where it is quite far from the resonant region. In the absence of DM resonant annihilation and other annihilation contributions that being not very significant, the fermionic DM yield is higher during the freeze-out, compared to that in scenario-I. 
Here, the fermionic DM relic density is over-abundant ($\sim 200$ times the observed relic density), when only the individual fermionic DM contribution is taken. 
The scalar dark matter here also stays far away from the resonant region, and $H$ mediated t-channel annihilation plays a major role. 
Non-resonant annihilation of the individual scalar DM leads to a relic density of under-abundant values. 
This is reflected in Fig.~\ref{fig:CaseII-200GeV} (left) where even $Y_{\phi_3}$ values at
freeze-out is smaller than $Y_{\chi_1}$ which is completely opposite to what is observed in scenario I. 

When exploring the coupled DM scenario for this reference point, again the yield hierarchy reverses to set $Y_{\phi_3}$ at higher values than the $Y_{\chi_1}$ at the freeze-out. 
This is a very significant phenomenon where one can observe the scalar and fermion DM densities to flip when single-component DM and coupled DM scenarios are considered. This is due to two reasons: the first reason is that the scalar DM gets boosted in the coupled DM set up, and thus annihilation becomes less efficient. Hence the relic density increases and therefore the scalar DM yield again becomes larger. 
The second reason is that the fermionic DM annihilates to scalar DM, reducing the fermionic relic density, which makes the yield flip more apparent. 
We find that, even if the fermionic DM relic comes down in the coupled case from the severely over-abundant value in the single DM scenario, it is not enough to become underabundant. Therefore total relic density in the coupled case remains over abundant for this benchmark point, which we discuss next.

Now we look at the relative importance of two dark matter components in the total relic density. Total relic abundance of the two-component DM scenario can be evaluated directly from the freeze-out plot using,
$$\Omega~h^2=\Omega_{\phi_3}h^2+\Omega_{\chi_1}h^2$$, 
where $\Omega_{\phi_3}h^2$ and $\Omega_{\chi_1}h^2$ are computed as:
\begin{eqnarray}
    \Omega_{\phi_3} h^2~=~\frac{m_{\phi_3}~s_0~Y_{\phi_3}(\infty)}{\rho_c/h^2} \hspace{0.7cm} 
    \Omega_{\chi_1} h^2=\frac{m_{\chi_1}~s_0~Y_{\chi_1}(\infty)}{\rho_c/h^2} \label{eq:4.14}
\end{eqnarray}
where $s_0=2890~{\rm cm^{-3}}$ is the current entropy density of the Universe and $\rho_c/h^2~=~1.05\times10^{-5}~{\rm GeV/cm^3}$ is the critical density. The asymptotic value of $Y$'s, both $Y_{\phi_3}(\infty)$ and $Y_{\chi_1}(\infty)$, is the value computed at the freeze-out point that remains constant for individual distribution of $Y$ till date. \\

\begin{table}[h!]
    \centering
    \begin{tabular}{|c|c|c|c|c|}
    \hline
     $m_{\chi_1}$ (GeV) & $m_{\phi_3}$ (MeV) & $\Omega_{\chi_1} h^2$ & $\Omega_{\phi_3} h^2$ & $\Omega h^2$ \\ [1ex]
      \hline \hline
       $45$ & $20$ & $6.5 \times 10^{-4} $ &  $1.4 \times 10^{-3}$ &  $2.05 \times 10^{-3}$ \\ [1ex]
       \hline 
       $45$ & $100$ & $1.2 \times 10^{-3}$ & 0.139 &  0.140 \\ [1ex]
       \hline 
       $200$ & $20$ & 5.71 & $2.08\times 10^{-3}$ & 5.712 \\ [1ex]
       \hline
       $200$ & $100$ & 5.88 & 0.167  & 6.047 \\ [1ex]
       \hline 
    \end{tabular}
    \caption{{The dark matter relic densities in a coupled two-component dark matter scenario, showing the relative contributions of the scalar and fermionic dark matters to the total relic density for different dark matter masses.}}
    \label{tab:13}
\end{table}

To provide a quantitative measure, the results of Fig.~\ref{fig:CaseI-65GeV} and Fig.~\ref{fig:CaseII-200GeV} are also shown numerically with variation of the relic density with the DM mass. In Table~\ref{tab:13}, we tabulate the relic abundance for both the dark matter candidates in the two-component coupled DM scenario, for the different masses of the fermionic and scalar DM. Their individual contributions to the total relic along with the total relic density are shown there. From these values, it can be concluded that when $\chi_1$ is around 45-50~GeV we get underabundant relic density if the scalar DM mass is below 100 MeV.

\begin{figure}[h!]
\centering
\begin{subfigure}{0.5\textwidth}
  \centering
  \includegraphics[width=0.95\linewidth]{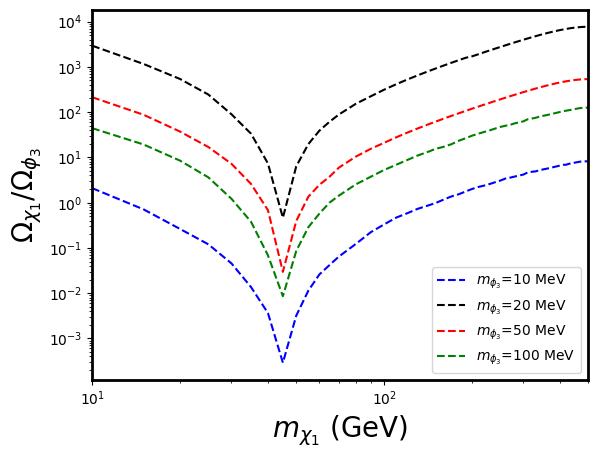}
  \label{fig:a}
\end{subfigure}%
\begin{subfigure}{.5\textwidth}
  \centering
  \includegraphics[width=0.95\linewidth]{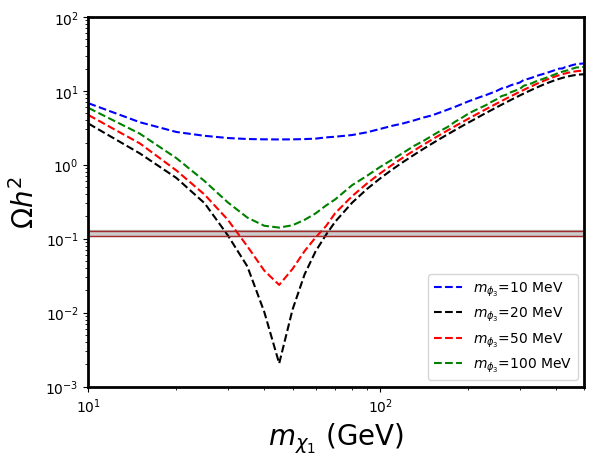}
  \label{fig:b}
\end{subfigure}
\caption{{Left: Ratio of the abundances $\Omega_{\chi_1}/\Omega_{\phi_3}$ as a function of fermion DM mass $m_{\chi_1}$ for four different scalar DM masses. Right: The total relic abundance $\Omega h^2 = \Omega_{\phi_3} h^2 + \Omega_{\chi_1} h^2$ as a function of fermion DM mass $m_{\chi_1}$. The grey solid band depicts the experimentally measured correct relic abundance within a $3 \sigma$ range.}}
\label{fig:RDvsMass}
\end{figure}

For Fig.~\ref{fig:RDvsMass}, we scan the fermionic DM mass $m_{\chi_1}$ from 10 GeV to 500 GeV and compute the total relic density adding the contributions from both the fermionic and the scalar DM components for a two-component coupled DM scenario. 
We fix four different values of the scalar DM mass $m_{\phi_3}$: 
10, 20, 50, and 100 MeV, and probe the relative importance of fermionic and scalar DM contributions in the total relic density.  
In the left plot, the ratio of the relic abundances $\Omega_{\chi_1} h^2$ and $\Omega_{\phi_3} h^2$ is shown as a function of the fermionic dark matter mass $m_{\chi_1}$. 
{Over the full fermionic DM range $m_{\chi_1}$ that is explored, different masses of the accompanying scalar DM lead to completely different predictions. 
For a scalar DM of 10 MeV, the fermionic DM relic is dominant throughout the range apart from the region of resonance, with the scalar DM contributing less than 1\% to 2\% of the total relic.} 
The corresponding total DM relic density is shown in Fig.~\ref{fig:RDvsMass} (right). Observationally underabundant or exact relic can be obtained if $m_{\chi_1} \sim (30-65)~$GeV. 
Note that in the single-component fermionic DM scenario, the underabundant region was very narrow, $\sim 40-50$~GeV. Hence, this is a very interesting modification in the fermion DM phenomenology that we have observed. This is due to the coupled nature of the two DM sectors: a new annihilation channel for the fermionic DM opens up in the form $\chi_1 \chi_1 \to \phi_3 \phi_3$. 
{It is important to note that both the $m_{\phi_3}=$10 and 100 MeV yield an over abundant relic density in the coupled scenario compared to the single scalar DM case, discussed in section \ref{sec:signle-DM}. 
The dashed green line for $m_{\phi_3}=$ 100 MeV scenario lies just above the correct relic band when $m_{\chi_1}=$ 45 GeV. But the blue dashed line for $m_{\phi_3}=$ 10 MeV case shows over abundant dark matter throughout the whole range of $m_{\chi_1}$.
This is the characteristic feature of the coupled scenario: the process $\chi_1 \chi_1 \to \phi_3 \phi_3$ enhances the scalar DM abundance. As a result, the resonant condition $m_{\phi_3} \sim m_H/2$ becomes less effective leading to an increase in relic density. Therefore from Fig. \ref{fig:RDvsMass}, it can be concluded that, in the coupled scenario, the underabundant or exact relic can be obtained if $20 ~\rm MeV \leq m_{\phi_3} <100~\rm MeV$.}

When the scalar DM mass is 50 MeV, its relic density increases to gradually become the dominant component of the total DM relic. 
The scalar DM contribution to the total relic can reach up to 
$>80$\% of the total relic density in this case for fermionic DM mass around $\sim 45$~GeV. This mass region is also very favorable from the total allowed DM relic density, which is shown in Fig.~\ref{fig:RDvsMass}(Right).
Scalar DM mass can be further increased to 100 MeV, and at this value, it starts to become the
dominant contributor to the dark matter relic. 
For all the fermionic DM masses explored here, the fermionic contribution remains dominant around (80-90)\% for $m_{\chi_1} \sim 10~$GeV and then decreases gradually for the higher DM values with around  <1\% contribution at $m_{\chi_1}
\sim 45~$GeV (for $m_{\phi_3} \sim 100~$MeV). 
The fermionic contribution again starts to increase further after this dip region. 
This is a significant deviation from the case in which individual DM candidates are studied. 
Previously, in the single component scenario, for $m_{\chi_1} \ge 100$ GeV the fermionic relic was always larger than a 100 MeV scalar DM relic. 
This is because in the coupled scenario, the light scalar DM gets a boost from the annihilation of the fermionic DM which makes its annihilation even less efficient, resulting in higher relic. Moreover, the relic density further increases for heavier scalar DM. 

In summary, when the total DM relic is either underabundant or exactly compatible with the experimental measurements, the fermionic DM contribution always dominates the scalar one. 
Obviously there is a parameter region where the scalar DM contribution is much higher than the fermionic counterpart. But in that scenario, total relic density is always over abundant i.e. ruled out by the experimental measurement.

\begin{figure}[!htb]
    \centering
    \includegraphics[scale=0.5]{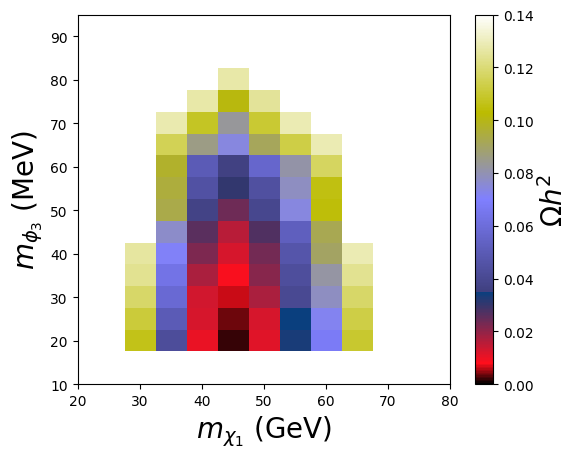}
    \caption{{The allowed region in the $m_{\chi_1} - m_{\phi_3}$ plane with under-abundant to exact relic abundance in the two-component coupled scenario. The $3\sigma$ upper limit on the dark matter relic density is placed as $\Omega h^2 \lesssim 0.13$ following the result by the Planck experiment. The white region is ruled out due to over-abundance of the relic density.}}
    \label{fig:mphi3-mchi1-relic}
\end{figure}

{
To further quantify the effect of boost on the DM relic density, we display the points with correct relic abundance in the $m_{\phi_3}-m_{\chi_1}$ plane corresponding to the coupled scenario. 
We perform a grid scan for $m_{\phi_3}$ in the range [10 MeV - 100 MeV] with a step size of 5 MeV, while $m_{\chi_1}$ is varied within the range [10 GeV - 100 GeV] with a step size of 5 GeV. 
The total relic density is calculated by solving the CBE, and adding up the respective relic from both of the components. 
Figure~\ref{fig:mphi3-mchi1-relic} illustrates the total relic density for the coupled scenario, incorporating contributions from both the scalar and fermionic dark matter. 
The yellowish regions have a total relic density $\lesssim 0.13$ (the 3$\sigma$ upper limit of the measured value), defining the outer boundary of the region allowed in the $m_{\phi_3}-m_{\chi_1}$ plane. 
For example, the bin at $m_{\chi_1}=45$ GeV on the x-axis shows that at 
$m_{\phi_3}=20$ MeV, the total relic density is at its lowest value. 
Now as $m_{\phi_3}$ increases, the relic density also rises, eventually reaching its upper limit $\lesssim 0.13$ at 
$m_{\phi_3}=80$ MeV. 
The white region is ruled out because that parameter space produces an overabundant relic. 
}

\section{Boost Effects in Dark  Matter Phenomenology} 
\label{sec:detection-boost}

{ 
Dark Matter particles in our Milky Way galaxy are (mostly) considered to be non-relativistic today. 
Because of the small velocity, the light DM can only produce smaller nuclear/electron recoil energy at the direct detection experiments. On the contrary, light dark matter particles can be boosted through different astrophysical phenomena, e.g. 
scattering with cosmic Ray particles, kinematically boosted through the presence of a heavier DM component etc..  Then the recoil energy can be large, and the boosted dark matter has a better possibility of detection at the direct detection experiments. Therefore boosted DM is in general is worthy to explore, even as generic case without the details of two component DM. Moreover, in a multi-component DM scenario, what fraction of the DM is boosted, will depend on its energy profile. A quantitative analysis along this direction is beyond the scope of this paper, we will study the detection aspects of boosted dark matter in a follow up work. In this section, we discuss the relic density aspects (in the early universe) of one and two component DM benchmark scenarios discussed in the previous section, and show how much boosts are associated with the lighter scalar DM particle.

\paragraph{\underline {Relic Density with MeV scale Boosted DM:}}

{First we study the single component light scalar DM ($\phi_3$) case without explicitly specifying how it is going to be boosted. In this case we work in the absence of the other DM component, the vectorlike DM $\chi_1$. We assume that $\phi_3$ is boosted through any generic boosting mechanism in the early Universe, while all the particles are inside the thermal bath.}
In such a general case, the boost factor $\gamma$ for the DM is defined as,
\begin{eqnarray}
\gamma_{\phi_3} = \frac{1}{\sqrt{1-v_{\phi_3}^2}},
\end{eqnarray}
and the COM energy in the annihilation processes of $\phi_3$ is expressed as
\begin{eqnarray}
 s=~4m_{\phi_3}^2+ 4m_{\phi_3}^2v_{\phi_3}^2.
\end{eqnarray}
In Fig.~\ref{fig:boost}, we plot the relic density of a single component DM ($\phi_3$) as a function of 
its mass for different boost ($\gamma_{\phi_3}$).
In the presence of boost, the COM energy increases, as deciphered from the second term in the above expression. 
As a result, the range of the kinematic variable $s$ used to compute thermal averaged cross section also increases with the new COM energy. 
Therefore, new annihilation channels should potentially open up for $\phi_3$, even at the lower masses, leading to additional annihilation channels that increase the total annihilation cross section. 
Hence in general for a boosted DM, we expect the relic density to be lower than the no boost scenario.

Interestingly, inclusion of the boost does opposite in our case: shifts the resonance drop towards left. 
In Fig.\ref{fig:boost}, the no boost scenario $\gamma_{\phi_3}=1$ is represented by the blue line, while the dashed lines show the effects of boost on the DM relic density.
Apart from the sharp relic drop at resonant annihilation region at $m_{\phi_3}=m_H/2$, we get the broader region with under-abundant relic because the $t$-channel $\phi_3\phi_3\to HH$ process opens up.
Here no new H mediated resonant annihilation channel opens up as the maximal energy that can be reached through the boost effects will not kinematically allow any new SM particle to get annihilated to. 
At the same time effect of the resonant condition $m_{\phi_3}\sim m_H/2$ gets diluted, leading to less effective DM annihilation, i.e. higher relic density as observed in Fig.~\ref{fig:boost}. 
Modified resonant conditions get satisfied with smaller values of the DM mass $m_{\phi_3}$, so the resonance dip shifts to the left. 

 \begin{figure}
 \centering
 \includegraphics[width=0.5\linewidth]{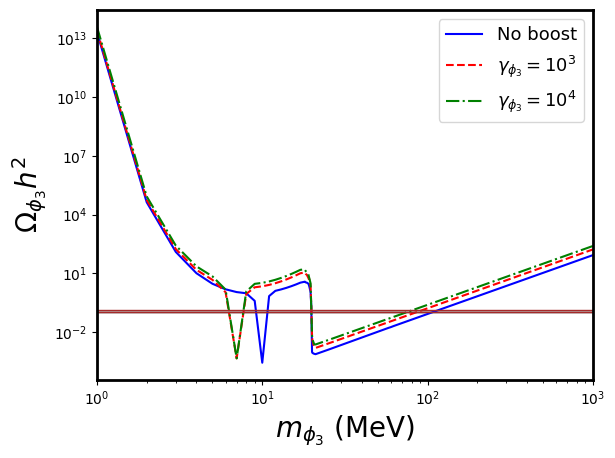}
\caption{{Variation of the relic density with its mass for single component scalar DM scenario showing the most general effects of a Lorentz boost. 
The blue line corresponds to the situation with no boost, $\gamma_{\phi_3}=1$. 
The red and green lines represent two cases, $\gamma_{\phi_3}= 10^3 \rm ~and~ 10^4$ respectively, without pointing out origin of the boost. 
The red band depicts the $3\sigma$ range of correct relic density, measured by the Planck experiment.}}
\label{fig:boost}
\end{figure}
}

\begin{figure}[h!]
\centering
  \includegraphics[width=0.41\linewidth]{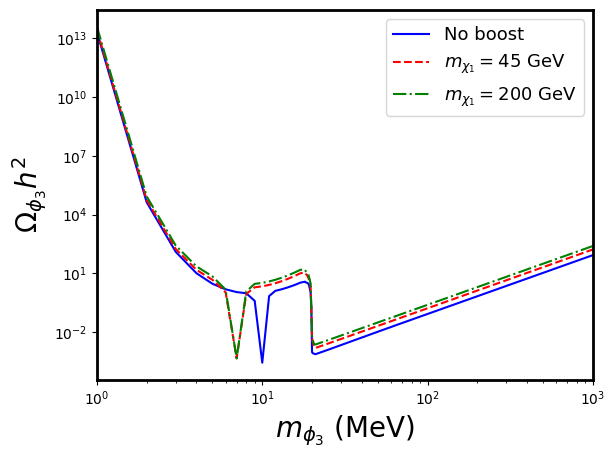}
 \includegraphics[width=0.45\linewidth]{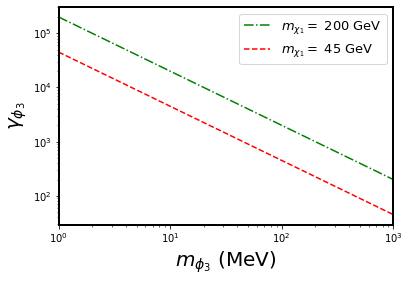}
\caption{{Left: The relic density of the boosted scalar DM as a function of its mass is shown for two benchmark values of the vectorlike DM in the two-component scenario. The blue solid line is the relic density of the scalar DM in the single component case with no boost. Right: The boost of the scalar DM and its variation with the DM mass is shown for two benchmark values of the vectorlike DM in the coupled scenario.
}}
\label{fig:boost1}
\end{figure}

Now let us discuss the boost effect in the two-component scenario. 
In this model set up, the annihilation process $\chi_1\chi_1\to \phi_3 \phi_3$ was dominant in the early universe. Hence, the light DM $\phi_3$ was boosted naturally from the annihilation of the Vectorlike DM ($\chi_1$). 
%
From the relativistic kinematic computation we obtain the boosted velocity of the scalar DM as:
\begin{eqnarray}\nonumber
v_{\phi_3}^2&=& 1-\frac{m_{\phi_3}^2}{m_{\chi_1}^2}~(1-v_{\chi_1}^2) \quad {\rm and~the~boost} \quad \gamma_{\phi_3} \sim \frac{m_{\chi_1}}{m_{\phi_3}}
 \label{eq:5.2}
\end{eqnarray}
The COM energy in the annihilation of the scalar DM is expressed as
\begin{eqnarray}
   s=~4m_{\phi_3}^2+4m_{\phi_3}^2~\big(1-\frac{m_{\phi_3}^2}{m_{\chi_1}^2}~(1-v_{\chi_1}^2)\big)  \label{eq:5.4}
\end{eqnarray}
Note that, here the boosted COM energy $s$ is now a function of $m_{\phi_3}$ and $m_{\chi_1}$.
We choose $v_{\chi_1}=~220~{\rm km/s}$ according to the Maxwell-Boltzmann distribution 
of the DM velocity distribution in Standard Halo Model (SHM) \cite{Evans:2005tn,Zemp:2008gw}.
In Fig.\ref{fig:boost1} (Left) we show the DM relic density as a function of the scalar DM mass for two benchmark masses of the 
vectorlike dark matter. The blue line shows the no boost scenario, as before. Here the scalar and the vectorlike dark matter 
together satisfy the total DM density as discussed in the previous section or in other words, the scalar DM satisfies only a fraction 
of the total DM density. When the vectorlike DM mass is 45 GeV, the total relic is satisfied. We also show in Fig.\ref{fig:boost1} (Right) that a significant amount of 
boost can be achieved in our benchmark cases and the amount of boost increases 
with the mass of the vectorlike DM.
In Fig.\ref{fig:boost1} (Left), the boost changes at every point of the plot as it depends on the masses of the scalar 
and the vectorlike dark matter. Thus, even if Fig.\ref{fig:boost1} (Left) and Fig.\ref{fig:boost} look almost similar, 
the boost of the scalar DM is not similar in the two cases. In the two-component model,  
correct relic density is achievable if the vectorlike DM mass is in the window 30-65 GeV with the scalar DM mass lying in the range of 20-80 MeV. 

For these masses, the boost factor of the scalar around $\mathcal{O}({10^3})$. Note that this is only true for the 
particular benchmark points that we choose in this model. A slight variation in the couplings of the vectorlike DM 
changes these limits. From Table \ref{tab:13} in the previous section and Fig.\ref{fig:boost1}, we find that for 
the scalar DM mass around 20 MeV and vectorlike DM mass $\sim 45$ GeV the relic density is exact or underabundant, and 
a significant boost is also achieved. Even higher boost is achievable for larger masses of the vectorlike DM, 
however, this tends to make the total relic to be overabundant in the particular benchmark points of the model parameters that we have chosen\footnote{We have also checked that the result remains almost unchanged if we vary $x$ in the range 30 to 100. Thus we keep  $x\simeq 20$ throughout the calculation.}.
%
\paragraph{\underline {Detection Prospects of MeV scale Boosted DM:}}
For the scalar DM, the strongest constraints are by XENON1T $^8$B \cite{XENON:2020gfr} 
and PandaX-4T $^8$B \cite{PandaX:2022aac} in the limit 4-10 GeV.
Other experiments such as XENON1T \cite{XENON:2018voc}, XENONnT \cite{XENONCollaboration:2023orw}, PandaX-4T \cite{PandaX-4T:2021bab} 
impose constraints on the spin-independent cross section when mass of the dark matter is more than 10 GeV.
However, the strongest limit comes from the LZ \cite{LUX-ZEPLIN:2022xrq} experiment.
dark matter direct detection searches become challenging in the lower mass region because the threshold for nuclear recoil energy is very low. Hence in the lower mass region, the strong limit comes from the electron scattering experiments such as 
DarkSide \cite{DarkSide:2018bpj}, PICO-60 \cite{PICO:2017tgi,PICO:2019vsc} and CRESST \cite{CRESST:2019jnq} and a series of other projects.
In Earth-based direct detection experiments, the dark matter particle interacts with 
either the nucleon or the electron of the atom. 
After the collision, the electron or the nucleon is recoiled with the 
energy transferred from the incoming dark matter particles. 
This recoil energy is then detected and measured by the experiments. 
If the mass of the DM particle is small (sub-GeV) it produces very small recoil energy, if not sufficiently boosted. 
On the other hand, if the DM travels with a large amount of kinetic energy, small mass DM can transfer large amount of energy to the recoil   
electron ($\phi_3 e^- \rightarrow \phi_3 e^-$) or nucleon ($\phi_3 n \rightarrow \phi_3 n $). Detection of these energetic light DM particles can be performed in various Dark Matter and Neutrino Experiments as illustrated in \cite{
Kim:2020ipj, Elor:2021swj, Ema:2018bih}.

The existing limits on the scalar DM can alter significantly if the particle is boosted. 
We discuss the effect of boost on the relic density of scalar DM  $\phi_3$ and it is shown that in the two-component scenario, the MeV scale dark matter that satisfies the total relic density 
together with the VLDM, is $\sim 10$ MeV. For MeV scale DM in this range, sufficient amount of boost can be obtained which can overcome the detector thresholds. There are challenges in the detection of the boosted dark matter in direct detection experiments.
First of all, the flux of the boosted DM particle is small and nearly mono-energetic \citep{Agashe:2015xkj}. Hence, for boosted dark matter, large volume detectors are preferred. However, the detection of the vectorlike DM ($\chi_1$) will be similar to the conventional WIMP direct detection studies. Nevertheless, as already mentioned, we will study the detection prospect of the boosted MeV scalar dark matter in detail in an upcoming work.

\section{Summary and Discussions}

In summary, we propose a low-mass scalar DM and a comparatively heavy-mass fermionic DM candidate in a modified $\nu \rm 2HDM$ scenario and discuss the DM phenomenology in the context of relic density. 
In the $\nu \rm 2HDM$ discussed here, there is one light CP even neutral scalar ($\approx 20~{\rm MeV}$) in addition to right-handed (RH) neutrinos $N_{R_i}$. One gauge singlet neutral scalar $\phi_3$ is added to the model which is stabilized to be the DM under the $Z_2^{DM}$ symmetry.The model is further extended with one vectorlike doublet ($N$) and one vectorlike singlet ($\chi$). 
Mixing of the vectorlike doublet ($N$) and vectorlike singlet ($\chi$) gives the second DM candidate of our model i.e. the fermionic DM $\chi_1$. 
The Yukawa sector of the $\rm 2HDM$ contains the interaction of the SM fermions with the Higgs bosons.
Along with other constraints, the oblique parameters constrain the model tightly because of the light scalar $H$, but the addition of vectorlike doublet and singlet 
allow for more favorable parameter space. 
After imposing other constraints such as Higgs and $Z$ boson invisible widths, stability of the potential etc, we choose the benchmark points from the allowed region of the parameter space.

In this model, both DM candidates annihilate to different SM particles, aided by s-channel resonant processes, along with some t-channel contributions. 
The Higgs-mediated channels (also the CP even scalar $H$) play a dominant role for both DM candidates in this scenario. 
Boltzmann equations are constructed to assess the thermal evolution of both DM candidates: first individually for them and then in a coupled DM scenario. 
To construct the Boltzmann equation, thermal averaged cross sections are used as inputs. 
Thermal averaging of a cross section is re-explained in both the relativistic and non-relativistic regimes.  
In the process of thermal averaging of the cross section, the limits on the mandelstam parameter $s$ is found to be important in distinguishing between relativistic and non-relativistic approaches.  
First we analyze the freeze-out of the scalar and fermionic DM in the uncoupled scenario. 
We find that the fermionic DM is very restricted while a light DM is somewhat favorable from the context of the relic density. 
However, the light-scalar DM suffers from strong constraints from detection due to its inability to recoil an electron or nucleus in the detector material. 
Then we consider the coupled scenario, where we have solved the
CBE (Coupled Boltzmann Equation) for the DM particles. 
The cross-term (interaction between the two DM sectors) in the CBE allows for additional annihilation channels for the fermion DM, and this is reflected in the delayed freeze-out of the fermion DM.  
On the other hand, for the scalar DM the s-channel resonant condition gets diluted, leading to less effective DM annihilation. 
Scalar DM relic density is enhanced for a mass range of 
$\sim [20-80]$ MeV compared to the individual scenario. Therefore, we find that the DM candidates satisfy the correct relic abundance for the $m_{\chi_1}\approx [30-65]~{\rm GeV}$ and $m_{\phi_3}$ values around 
$20-80$~MeV in the coupled DM scenario.

The choice of a light MeV scalar DM and a heavy fermionic DM, is crucial to have a boosted DM in the two-component DM scenario. 
The fermionic DM is said to be non-relativistic with an rms velocity $\approx 220~{\rm Km/s}$ in the galactic halo. 
However, because of the heavy mass, it can annihilate to $\phi_3$ (based on our study), and the conservation of energy allows $\phi_3$ to have more kinetic energy.
We observe that the boost received from the $\chi_1\chi_1\to \phi_3\phi_3$ annihilation is equivalent to a relativistic factor $\gamma \sim [10^3 - 10^4]$ 
in our benchmark scenarios. Even though the boosted light DM relic density is almost independent of how it is boosted, we successfully
find that in a two-component DM model which also satisfies the other phenomenological constraints. 
In the region where the relic density is underabundant, the contribution of the scalar DM ranges from 
$\sim 30 - 80 \%$ to the total DM with a significant boost.  
%
This boosted DM has special importance in the DM direct detection study, where a DM candidate hits a detector-level particle (electron or nucleon) and generates measurable recoil energy for the electron and nucleon.  
Therefore, a more energetic DM candidate (DM with boost) will produce more recoil energy, and thus the possibility of its detection will be enhanced. In our future work, we discuss the direct detection prospects of MeV scale light scalar DM, boosted from the heavier fermionic DM. Even as the relic from the boosted DM in our model can go up to almost 80\% of the total DM, cosmological constraints can potentially restrict the boosted contribution to some intermediate values. As both the DM candidates in the two-component model interact with the SM and also amongst each other, the future study on the dark matter detection prospect will assess the viability of the model.


\section*{Acknowledgments}
The authors would like to thank Divya Sachdeva and Subhendra Mohanty for useful discussions. The work of NK is supported by the Department of Science and Technology, Government of India under the SRG grant, Grant Agreement Number SRG/2022/000363 and CRG grant with Grant Agreement Number CRG/2022/004120. The work of AB and AC is funded by the Department of Science and Technology, Government of India, under Grant No. IFA18-PH 224 (INSPIRE Faculty Award). SS thanks Vivekananda Centre for Research (VCR) for providing the research facilities.

\newpage

\appendix
\section{Dark Matter Annihilation cross sections}
\label{process}



\subsection{Scalar DM annihilation cross sections} \label{sec:A.1}

\begin{itemize}
\item {\fbox{$\phi_3\phi_3\rightarrow f\bar{f}$ }} 
\begin{eqnarray} \nonumber
    \sigma(s)_{\phi_3\phi_3\rightarrow f\bar{f}} &=&\frac{1}{16\pi s}\frac{\sqrt{s-4m_f^2}~(s-4m_f^2)}{\sqrt{s-4m_{\phi_3}^2}}\Bigg[\frac{\lambda_{\phi_3\phi_3h}^2~\lambda_{ffh}^2}{(s-m_h^2)^2}\\ \nonumber &+&
    \frac{\lambda_{\phi_3\phi_3H}^2~\lambda_{ffH}^2}{(s-m_H^2)^2}\\ &+&  \frac{2\lambda_{\phi_3\phi_3h}~\lambda_{ffh}\lambda_{\phi_3\phi_3H}~\lambda_{ffH}}{(s-m_h^2)(s-m_H^2)}\Bigg] \label{eq:A:2}
\end{eqnarray} 
\par

\item {\fbox{$\phi_3\phi_3\rightarrow N_R\;\nu_L$}}
\begin{eqnarray} \nonumber
    \sigma(s)_{\phi_3\phi_3\to N_R\nu_L}&=&\frac{1}{16\pi s}\frac{\sqrt{s-m_{N_R}^2}~(s-m_{N_R}^2)}{\sqrt{s-4m_{\phi_3}^2}}\Bigg[\frac{\lambda_{\phi_3\phi_3h}^2~\lambda_{h\nu_L N_R}^2}{(s-m_h^2)^2}\\ \nonumber &+&\frac{\lambda_{\phi_3\phi_3H}^2~\lambda_{H\nu_L N_R}^2}{(s-m_H^2)^2}\\ &+&\frac{2~\lambda_{\phi_3\phi_3h}~\lambda_{\phi_3\phi_3H}~\lambda_{h\nu_L N_R}~\lambda_{H\nu_L N_R}}{(s-m_h^2)(s-m_H^2)} \Bigg] \label{eq:A.4}
\end{eqnarray}

\item {\fbox{$\phi_3\phi_3\;\rightarrow\; hh$}}
\begin{eqnarray} \nonumber
  \sigma(s)_{\phi_3\phi_3\to hh}&=&\frac{(\lambda_{\phi_3\phi_3h})^4}{\pi s}\frac{\sqrt{s-4m_h^2}}{\sqrt{s-4m_{\phi_3}^2}}\frac{1}{4m_h^4-16m_h^2m_{\phi_3}^2+4sm_{\phi_3}^2}+\frac{1}{16\pi s}\frac{\sqrt{s-4m_h^2}}{\sqrt{s-4m_{\phi_3}^2}}\Bigg[(\lambda_{\phi_3\phi_3}^{hh})^2\\ \nonumber &+& \frac{(\lambda_{\phi_3\phi_3h}~\lambda_{hhh})^2}{(s-m_h^2)^2}+ \frac{(\lambda_{\phi_3\phi_3H}~\lambda_{hhH})^2}{(s-m_H^2)^2}+\frac{2\lambda_{\phi_3\phi_3}^{hh}\lambda_{\phi_3\phi_3h}\lambda_{hhh}}{(s-m_h^2)}\\\nonumber &+& \frac{2\lambda_{\phi_3\phi_3}^{hh}\lambda_{\phi_3\phi_3H}\lambda_{hhH}}{(s-m_H^2)}+\frac{2\lambda_{\phi_3\phi_3h}\lambda_{hhh}~\lambda_{\phi_3\phi_3H}~\lambda_{hhH}}{(s-m_h^2)(s-m_H^2)}\Bigg]\\&+&  \log\frac{2m_h^2-s+\sqrt{s-4m_{\phi_3}^2}\sqrt{s-4m_h^2}}{2m_h^2-s-\sqrt{s-4m_{\phi_3}^2}~\sqrt{s-4m_h^2}} \Bigg[\frac{\lambda_{\phi_3\phi_3}^{hh}(\lambda_{\phi_3\phi_3h})^2}{4\pi s(s-4m_{\phi_3}^2)}\\ \nonumber &+&\frac{(\lambda_{\phi_3\phi_3h})^3\lambda_{hhh}}{4\pi s(s-4m_{\phi_3}^2)(s-m_h^2)}+\frac{(\lambda_{\phi_3\phi_3h})^2\lambda_{\phi_3\phi_3H}\lambda_{hhH}}{4\pi s(s-4m_{\phi_3}^2)(s-m_H^2)}\Bigg]
\label{eq:A.6}
\end{eqnarray}
\item {\fbox{$\phi_3\phi_3\;\rightarrow\; HH$}}
\begin{eqnarray} \nonumber
  \sigma(s)_{\phi_3\phi_3\to HH}&=&\frac{(\lambda_{\phi_3\phi_3H})^4}{\pi s}\frac{\sqrt{s-4m_H^2}}{\sqrt{s-4m_{\phi_3}^2}}\frac{1}{4m_H^4-16m_H^2m_{\phi_3}^2
  +4sm_{\phi_3}^2}+\frac{1}{16\pi s}\frac{\sqrt{s-4m_H^2}}{\sqrt{s-4m_{\phi_3}^2}}\Bigg[(\lambda_{\phi_3\phi_3}^{HH})^2 \\ \nonumber
  &+& \frac{(\lambda_{\phi_3\phi_3H}\lambda_{HHH})^2}{(s-m_H^2)^2} 
  + \frac{(\lambda_{\phi_3\phi_3h}~\lambda_{hHH})^2}{(s-m_h^2)^2}+\frac{2\lambda_{\phi_3\phi_3}^{HH}\lambda_{\phi_3\phi_3H}~\lambda_{HHH}}{(s-m_H^2)} \\ \nonumber  
  &+&\frac{2\lambda_{\phi_3\phi_3}^{HH}\lambda_{\phi_3\phi_3h}\lambda_{hHH}}{(s-m_h^2)}+\frac{2\lambda_{\phi_3\phi_3h}\lambda_{HHH}\lambda_{\phi_3\phi_3H}\lambda_{hHH}}{(s-m_h^2)(s-m_H^2)}\Bigg] \\ 
  &+&  \log\frac{2m_H^2-s+\sqrt{s-4m_{\phi_3}^2}\sqrt{s-4m_H^2}}{2m_H^2-s-\sqrt{s-4m_{\phi_3}^2}\sqrt{s-4m_H^2}} \Bigg[\frac{\lambda_{\phi_3\phi_3}^{HH}(\lambda_{\phi_3\phi_3H})^2}{4\pi s(s-4m_{\phi_3}^2)} \\ \nonumber 
  &+&\frac{(\lambda_{\phi_3\phi_3H})^3\lambda_{HHH}}{4\pi s(s-4m_{\phi_3}^2)(s-m_H^2)}+\frac{(\lambda_{\phi_3\phi_3H})^2\lambda_{\phi_3\phi_3h}\lambda_{hHH}}{4\pi s(s-4m_{\phi_3}^2)(s-m_h^2)}\Bigg] \label{eq:A.8}
\end{eqnarray}
\par
\item {\fbox{$\phi_3\phi_3\;\rightarrow\;hH$}}
\begin{eqnarray} \nonumber
    \sigma(s)_{\phi_3\phi_3\to hH}&=&\frac{1}{16\pi s}\frac{\sqrt{s-(m_h+m_H)^2}}{\sqrt{s-4m_{\phi_3}^2}}\Bigg[~\frac{(\lambda_{\phi_3\phi_3h}~\lambda_{hhH})^2}{(s-m_h^2)^2}+\frac{(\lambda_{\phi_3\phi_3H}~\lambda_{hHH})^2}{(s-m_H^2)^2}+(\lambda_{\phi_3\phi_3}^{hH})^2\\&+& \frac{2\lambda_{\phi_3\phi_3h}~\lambda_{\phi_3\phi_3H}~\lambda_{hhH}~\lambda_{hHH}}{(s-m_h^2)(s-m_H^2)}+\frac{2\lambda_{\phi_3\phi_3}^{hH}~\lambda_{\phi_3\phi_3h}~\lambda_{hhH}}{(s-m_h^2)}+\frac{2\lambda_{\phi_3\phi_3}^{hH}~\lambda_{\phi_3\phi_3H}~\lambda_{hHH}}{(s-m_H^2)}~\Bigg] \nonumber \\\label{eq:A.10}
\end{eqnarray}
\end{itemize}

\subsection{Fermionic DM annihilation cross sections} \label{sec:A.2}

\begin{itemize}
\item {\fbox{$\chi_1\;\chi_1\rightarrow\;\phi_3\;\phi_3$}}
\begin{eqnarray} 
        \sigma_{\chi_1\chi_1\to\phi_3\phi_3}&=&\frac{1}{16\pi }\;\frac{\sqrt{s-4m_{\phi_3}^2}}{\sqrt{s-4m_{\chi_1}^2}}\Bigg[\frac{(\lambda_{\chi_1 \chi_1 h}~\lambda_{\phi_3 \phi_3 h})^2}{(s-m_h^2)^2} + \frac{(\lambda_{\chi_1 \chi_1 H}~\lambda_{\phi_3 \phi_3 H})^2}{(s-m_H^2)^2}\nonumber \\  &+& \frac{\lambda_{\chi_1 \chi_1 h}~\lambda_{\phi_3 \phi_3 h}~\lambda_{\chi_1\chi_1 H}~\lambda_{\phi_3\phi_3 H}}{(s-m_h^2)(s-m_H^2)} \Bigg] \label{eq:A.6}
\end{eqnarray}

\item  {\fbox{$\chi_1\chi_1\rightarrow f\bar{f}$}}
\begin{eqnarray} 
    \sigma(s)_{\chi_1\chi_1\to f\bar{f}}&=&\frac{(\lambda_{\chi_1\chi_1h~\lambda_{hf\bar{f}}})^2~(s-4m_f^2)^{3/2}}{4\pi\big((s-m_h^2)^2+{\Gamma_h^2m_h^2}\big)\sqrt{s-4m_{\chi_1}^2}}~+\frac{\lambda_{\chi_1\chi_1 Z}^2\sqrt{s-4m_f^2}}{4\pi s\sqrt{s-4m_{\chi_1}^2}~\big((s-M_Z^2)^2+ \Gamma_Z^2M_Z^2 \big)} \nonumber \\ 
     &\times & \Bigg[\frac{1}{3}(c_A^2+c_V^2)(4s^2+16m_f^2m_\chi^2 -4s(m_\chi^2+m_f^2)) \nonumber \\  
&+& 2m_\chi^2(c_A^2+c_V^2)(s-2m_f^2)+2m_f^2 s (c_A^2-c_V^2)\Bigg]
    \label{eq:A.7}
\end{eqnarray}
\item  {\fbox{$\chi_1\chi_1\rightarrow N_R\nu_L$}}
\begin{eqnarray}  \nonumber
    \sigma(s)_{\chi_1\chi_1\to N_R\nu_L} &=&\frac{1}{16\pi} \frac{\sqrt{s-m_{N_R}^2} (s-m_{N_R}^2)}{\sqrt{s-4m_{\chi_1}^2}} \Bigg[ \frac{(\lambda_{\chi_1\chi_1H}~\lambda_{HN_R\nu_L})^2}{(s-m_H^2)^2} + \frac{(\lambda_{\chi_1\chi_1h} \lambda_{hN_R\nu_L})^2}{(s-m_h^2)^2} \\  &+& \frac{\lambda_{\chi_1\chi_1H}~\lambda_{\chi_1\chi_1h}~\lambda_{HN_R\nu_L}~\lambda_{hN_R\nu_L}}{(s-m_h^2)~(s-m_H^2)}\Bigg] \label{eq:A.8}
\end{eqnarray}
\item  {\fbox{$\chi_1~\chi_1\;\rightarrow\;h~h~$(t channel)}}
\begin{eqnarray}
    \sigma(s)_{\chi_1\chi_1\to hh} &=&\frac{\lambda_{\chi_1\chi_1h}^4}{\pi (s-4m_{\chi_1}^2)}~\Bigg[\frac{8m_{\chi_1}^2\sqrt{s-4m_{\chi_1}^2}\sqrt{s-4m_h^2}}{4m_h^4-16m_h^2m_{\chi_1}^2+4sm_{\chi_1}^2}-\log\frac{2m_h^2-s-\sqrt{s-4m_{\chi_1}^2}\sqrt{s-4m_h^2}}{2m_h^2-s+\sqrt{s-4m_{\chi_1}^2}\sqrt{s-4m_h^2}}\Bigg] \nonumber \\  &+& \frac{\lambda_{\chi_1\chi_2h}^4}{\pi (s-4m_{\chi_1}^2)} \Bigg[~\frac{4(m_{\chi_1}^2+m_{\chi_2}^2)\sqrt{s-4m_{\chi_1}^2}\sqrt{s-4m_h^2}}{(2m_{\chi_1}^2-2m_{\chi_2}^2+2m_h^2-s)^2-(s-4m_{\chi_1}^2)(s-4m_h^2)} \nonumber \\  &-&\log\frac{2m_{\chi_1}^2+2m_h^2-2m_{\chi_2}^2-s-\sqrt{s-4m_{\chi_1}^2}\sqrt{s-4m_h^2}}{2m_{\chi_1}^2+2m_h^2-2m_{\chi_2}^2-s+\sqrt{s-4m_{\chi_1}^2}\sqrt{s-4m_h^2}}\Bigg]+\frac{\lambda_{\chi_1\chi_1h}^2\lambda_{hhh}^2}{8\pi (s-m_h^2)^2}\frac{\sqrt{s-4m_h^2}}{\sqrt{s-4m_{\chi_1}^2}} \nonumber \\  &+& \frac{\lambda_{\chi_1\chi_1h}^2~\lambda_{\chi_1\chi_2h}^2}{\pi (s-4m_{\chi_1}^2)}~\Bigg[\log\frac{2m_{\chi_1}^2+2m_h^2-2m_{\chi_2}^2-s+\sqrt{s-4m_{\chi_1}^2}\sqrt{s-4m_h^2}}{2m_{\chi_1}^2+2m_h^2-2m_{\chi_2}^2-s-\sqrt{s-4m_{\chi_1}^2}\sqrt{s-4m_h^2}}+\frac{2m_{\chi_1}^2}{(m_{\chi_1}^2-m_{\chi_2}^2)} \nonumber \\   & \times & \Bigg[\log\frac{2m_h^2-s+\sqrt{s-4m_{\chi_1}^2}\sqrt{s-4m_h^2}}{2m_h^2-s-\sqrt{s-4m_{\chi_1}^2}\sqrt{s-4m_h^2}}\nonumber \\  &-& \log\frac{2m_{\chi_1}^2+2m_h^2-2m_{\chi_2}^2-s+\sqrt{s-4m_{\chi_1}^2}\sqrt{s-4m_h^2}}{2m_{\chi_1}^2+2m_h^2-2m_{\chi_2}^2-s-\sqrt{s-4m_{\chi_1}^2}\sqrt{s-4m_h^2}}\Bigg]\Bigg] \nonumber \\  &+& \frac{\lambda_{\chi_1\chi_1h}\lambda_{hhh}~m_{\chi_1}}{4\pi(s-m_h^2)(s-4m_{\chi_1}^2)}\Bigg[\lambda_{\chi_1\chi_1h}^2\log\frac{2m_h^2-s+\sqrt{s-4m_{\chi_1}^2}\sqrt{s-4m_h^2}}{2m_h^2-s-\sqrt{s-4m_{\chi_1}^2}\sqrt{s-4m_h^2}} \nonumber \\  &+& \lambda_{\chi_1\chi_2h}^2\log\frac{2m_{\chi_1}^2+2m_h^2-2m_{\chi_2}^2-s+\sqrt{s-4m_{\chi_1}^2}\sqrt{s-4m_h^2}}{2m_{\chi_1}^2+2m_h^2-2m_{\chi_2}^2-s-\sqrt{s-4m_{\chi_1}^2}\sqrt{s-4m_h^2}}\Bigg] \label{A.14}
\end{eqnarray}
\item  {\fbox{$\chi_1~\chi_1\;\rightarrow\;Z~Z~$}}
\begin{eqnarray} 
    \sigma(s)_{\chi_1\chi_1\to ZZ} &=&\frac{\sqrt{s-4m_Z^2}}{16\pi\sqrt{s-4m_{\chi_1}^2}}\Bigg[\frac{\lambda_{\chi_1\chi_1Z}^4(12m_Z^4+s^2-4sm_Z^2)}{m_Z^2(m_Z^4-4m_Z^2m_{\chi_1}^2+sm_{\chi_1}^2)} + \frac{(\lambda_{\chi_1\chi_1 h}~\lambda_{hZZ})^2(12m_Z^4+s^2-4sm_Z^2)}{(s-m_h^2)^2~4m_Z^4} \nonumber \\  &+& \frac{\lambda_{\chi_1\chi_2Z}^4(12m_Z^4+s^2-4sm_Z^2)}{m_Z^2\Big((2m_{\chi_1}^2+2m_h^2-2m_{\chi_2}^2-s)^2-(s-4m_{\chi_1}^2)(s-4m_Z^2)\Big)} \Bigg]\nonumber \\  &+&  \frac{1}{16\pi}\frac{\lambda_{\chi_1\chi_1 h}~\lambda_{hZZ}~m_{\chi_1}}{(s-4m_{\chi_1}^2)(s-m_h^2)m_Z^4} \Bigg[ \lambda_{\chi_1\chi_1Z}^2(12m_Z^4+s^2-4sm_Z^2)\nonumber \\ &\times &\log\frac{2m_Z^2-s+\sqrt{s-4m_Z^2}\sqrt{s-4m_{\chi_1}^2}}{2m_Z^2-s-\sqrt{s-4m_Z^2}\sqrt{s-4m_{\chi_1}^2}} \nonumber \\  &+& \lambda_{\chi_1\chi_2Z}^2(12m_Z^4+s^2-4sm_Z^2)\log\frac{2m_{\chi_1}^2+2m_Z^2-2m_{\chi_2}^2-s+\sqrt{s-4m_Z^2}\sqrt{s-4m_{\chi_1}^2}}{2m_{\chi_1}^2+2m_Z^2-2m_{\chi_2}^2-s-\sqrt{s-4m_Z^2}\sqrt{s-4m_{\chi_1}^2}} \Bigg]\nonumber \\  &+& \frac{\lambda_{\chi_1\chi_1Z}^2\lambda_{\chi_1\chi_2Z}^2}{16\pi}\frac{12m_Z^4+s^2-4sm_Z^2}{m_z^2(s-4m_{\chi_1}^2)(m_{\chi_1}^2-m_{\chi_2}^2)}\Bigg[\log\frac{2m_Z^2-s+\sqrt{s-4m_Z^2}\sqrt{s-4m_{\chi_1}^2}}{2m_Z^2-s-\sqrt{s-4m_Z^2}\sqrt{s-4m_{\chi_1}^2}} \nonumber \\  &-& \log\frac{2m_{\chi_1}^2+2m_Z^2-2m_{\chi_2}^2-s+\sqrt{s-4m_Z^2}\sqrt{s-4m_{\chi_1}^2}}{2m_{\chi_1}^2+2m_Z^2-2m_{\chi_2}^2-s-\sqrt{s-4m_Z^2}\sqrt{s-4m_{\chi_1}^2}}\Bigg] \label{A.15}
\end{eqnarray}

\item  {\fbox{$\chi_1~\chi_1\;\rightarrow\; W^+~W^-$}}
\begin{align} \nonumber
    \sigma(s)_{\chi_1\chi_1\to W^+W^-} 
 = \frac{\sqrt{s-4m_W^2}}{4\pi\sqrt{s-4m_{\chi_1}^2}}~\Bigg[\frac{(\lambda_{\chi_1\chi_1 h}~\lambda_{hW^+W^-})^2(12m_W^4+s^2-4sm_W^2)}{16m_W^2(s-m_h^2)^2} \\\nonumber 
 + \frac{\lambda_{\chi_1N_-W^+}^2\lambda_{\chi_1N_+W^-}^2~(12m_W^4+s^2-4sm_W^2)}{m_W^2(4m_{\chi_1}^4+4m_W^4+m_N^4-4m_{\chi_1}^2m_N^2-4m_W^2m_N^2-8m_{\chi_1}^2m_W^2+2m_N^2s)}\Bigg]  \\ \nonumber 
 + \frac{1}{16\pi}\frac{(\lambda_{\chi_1\chi_1 h}~\lambda_{hW^+W^-}~\lambda_{\chi_1N_-W^+}~\lambda_{\chi_1N_+W^-})~m_{\chi_1}~(12m_W^4+s^2-4sm_W^2)}{(s-4m_{\chi_1}^2)~(s-m_h^2)~4m_W^4}  \\
  \times \log\frac{2m_{\chi_1}^2+2m_W^2-m_N^2-s+\sqrt{s-4m_W^2}\sqrt{s-4m_{\chi_1}^2}}{2m_{\chi_1}^2+2m_W^2-m_N^2-s-\sqrt{s-4m_W^2}\sqrt{s-4m_{\chi_1}^2}} \label{A.16}
\end{align}

\end{itemize}



\bibliographystyle{apsrev}
\bibliography{ref}


\end{document}